\documentstyle[preprint,aps,epsf]{revtex}
\tighten

\newcommand{\lsim}{\lower .5ex\hbox{$\buildrel < \over {\sim}$}}

\preprint{\vbox{Submitted to Physical Review D \hfill }}  

\begin{document}


\title{The Effects of Correlations on Neutrino Opacities \\ in Nuclear Matter}
\author{Adam Burrows}
\address{Department of Astronomy, The University of Arizona, Tucson, AZ 85721
\\ e-mail: burrows@as.arizona.edu}
\author{R. F. Sawyer}
\address{Department of Physics, 
The University of California,
Santa Barbara, CA 93106
\\ e-mail: sawyer@sarek.physics.ucsb.edu}
\maketitle
\setcounter{page}{1}
\begin{abstract}

Including nucleon--nucleon correlations due to 
both Fermi statistics and nuclear forces, we have 
developed a general formalism for calculating
the neutral--current neutrino--nucleon scattering rates in
nuclear matter. 
We derive corrections to the dynamic structure factors
due to both density and spin correlations
and find that neutrino--nucleon scattering rates are suppressed
by large factors around and above nuclear density.  
Hence, in particular for the $\nu_\mu$ and $\nu_\tau$ neutrinos,
but also for the $\nu_e$ neutrinos,
supernova cores are more ``transparent" 
than previously thought.  The many--body
corrections increase with density, decrease with temperature,
and are roughly independent of incident neutrino energy.

In addition, we find that
the spectrum of energy transfers in neutrino scattering is
considerably broadened by the interactions in the medium. An
identifiable component of this broadening comes from the
absorption and emission of quanta of collective modes akin to the
Gamow--Teller and Giant Dipole resonances in nuclei (zero-sound;
spin sound), with \v{C}erenkov kinematics.

Under the assumption that both the charged--current and the 
neutral--current cross sections are decreased by many--body effects, 
we calculate a set of ad hoc protoneutron
star cooling models to gauge the potential importance of the new opacities
to the supernova itself.  While the early luminosities are
not altered, the luminosities after many hundreds of milliseconds to
seconds can be increased by factors that range from 10\% to 100\%.
Such enhancements may have a bearing on the efficacy of the
neutrino--driven supernova mechanism, the delay
to explosion, the energy of the explosion,
and the strength and relative role of convective overturn
at late times.  However, the actual consequences, if any, of these new
neutrino opacities remain to be determined.

\end{abstract}

\pacs{PACS numbers: 97.60.Bw, 97.60Jd, 25.30.Pt, 11.80.Jy, 26.50.+x, 05.60.+w, 11.80.Gw, 12.15.Mn}



\section{Introduction}

In core--collapse supernovae, neutrinos are arguably the engines of explosion 
\cite{bw,bhf,herant,jankaM,mw} and the best direct probe of its internal workings
\cite{hir,bion,bur1,bl}.  To understand the supernova phenomenon, a theorist 
requires knowledge of the equation of state of nuclear matter, stellar evolution,
general relativity, statistical physics, and the techniques of radiative 
transfer and hydrodynamics. However, the opacities and sources of neutrinos
of all the six known neutrino species take center stage in this context in which the neutrino 
``optical" depth down to the protoneutron star core varies from $10^3$ to $10^6$, 
neutrino luminosities can approach $10^{54}$ ergs s$^{-1}$, and photons,
the traditional agents of radiative energy transfer in astrophysics, 
are profoundly trapped.  This seems to be a unique role for neutrinos in the 
universe.  Since neutrinos are abundantly produced, yet are the most mobile
of a collapsed core's constituents, their microphysics determines 
the outcome of core collapse, neutron star and black hole formation,
nucleosynthesis at and beyond the iron peak \cite{ww,timmes}, early pulsar kinematics \cite{bh}, 
and, perhaps, the asymmetries observed in supernova debris clouds \cite{wooden}. 

In the neutrino--driven mechanism, neutrinos liberated from the core
heat matter in the inner stellar envelope (the outer protoneutron star envelope)
as they emerge.  If the total heating rate is sufficient, the envelope
becomes unstable and is ejected.   It has recently been shown \cite{bhf,herant,jankaM,lichten}
that neutrino--driven convection in this inner mantle can make this process 
more efficient (but see \cite{mezz2}). Hence, neutrinos mediate the energy transfer
from the core to the mantle and their luminosities and spectra are crucial
to the viablility and character of the explosion mechanism.  These in turn
are a function of the neutrino--matter opacities.  

While at low mass densities ($\rho < 10^{12}$ gm cm$^{-3}$) 
neutrino--matter cross sections are well--understood 
and characterized \cite{tusc,glas,wein,bru2}, 
at the higher densities achieved in core collapse the inter--particle 
spacings are smaller than the wavelength of the ambient neutrinos 
and many--body effects and particle--particle correlations must be
taken into account. Ion--ion correlations due to Coulomb interactions have been addressed 
in the supernova context ({\it cf.} \cite{horo}) and their effects, 
though interesting, have been shown to be small and transient \cite{mezzbru}.  
However, near and above nuclear densities ($\sim2.6\times10^{14}$ gm cm$^{-3}$)
neutrinos with energies below $\sim$1000 MeV ``see" many nucleons
and collective effects must be considered.  In fact, these effects
can radically alter the opacity of nuclear matter to neutrinos, since the nuclear
force introduces strong correlations near and above nuclear densities.

Many--body effects have been discussed before in the 
protoneutron star and supernova context 
({\it cf.}, \cite{iwa,gp,keiljr,jankak,sigl}), but the results
were either too approximate or of too--limited scope.   Pauli blocking
by final--state nucleons, a type of correlation, has been a component 
of supernova and protoneutron star thinking for some time \cite{bl,bru2,schn}, 
but the approximate treatments employed at high densities can be off by large factors (\S VI.B).
Furthermore, in the case of neutrino--nucleon scattering, the assumption
that the scattering is elastic \cite{lp} and that the nucleons are stationary
can be shown to result in errors in the energy transfer of as much as a factor of ten.
A consequence of this is that energy equilibration of $\nu_\mu$ and $\nu_\tau$
neutrinos is brought about not by the $\nu + \bar{\nu}\rightarrow e^+ + e^-$
process or neutrino--electron scattering, but by neutrino--nucleon scattering.
This result, one by-product of this study, has also been found  
by Hannestad \& Raffelt \cite{hr} and in the pioneering 
work of Reddy {\it et al.} \cite{reddy}.
The latter have conducted the most complete exploration of high--density
neutrino opacities to date, taking care to calculate them in a manner consistent 
with a reasonable nuclear equation of state (see also \cite{reddy2,reddy3}).
Some of our results and formalism recapitulate 
theirs.  However, Reddy {\it et al.} \cite{reddy}
neglected particle--particle and spin--spin correlations 
and focussed on the kinematic effects of nuclear interactions. 
We find that the effect of many--body correlations 
on neutrino--matter cross sections can be 
quite large.  Nuclear matter is more ``transparent" to 
neutrinos than we had heretofore imagined.

Neutrinos interact with nucleons via both charged--current and neutral--current
processes.  The charged--current processes predominate for the electron-type
neutrinos by a few factors, while the neutral current predominates for 
$\nu_\mu$ and $\nu_\tau$ neutrinos. The latter carry away more than 50\% of the 
binding energy of the neutron star.  We have found that at the high densities
achieved in collapse, both the charged--current and the neutral--current
cross sections are dramatically affected by nuclear correlation effects.
In this paper, we focus on neutral--current neutrino--nucleon 
scattering, both the Fermi and the Gamow--Teller
parts, and defer a discussion of the charged--current many--body effects
to a later paper.  Our formalism is good for any degree of nucleon degeneracy
and fully incorporates the effects of reaction kinetics, Pauli blocking,
and correlations due to interactions.  Energy transfers both to and from the nucleons
are consistently included.  One product of this paper is a formalism for 
calculating in full the dynamic structure functions for neutrino--nucleon processes. 

The effect of increased transparency at high densities 
on the neutrino--driven mechanism of core--collapse
supernovae is not yet clear.  Supernova theory is notorious for its 
mitigating feedbacks and false leads.  The early ($\lsim 500$ milliseconds)
neutrino luminosities and spectra depend upon the material in
the outer shocked mantle of the protoneutron star, where we think we
understand neutrino cross sections.  As a consequence, we don't
expect that our understanding of this early phase will be altered by the
new opacities.  Even if it could be shown that neutrino cross sections at the 
lower densities around the neutrinospheres and near the shock were in some
non--trivial sense different, increased transparency would lead to an
increased rate of collapse of the mantle, which, in turn, would
lead to higher densities, which would partially quench the effect.
The corrections to the scattering rates we address in this paper are more
relevant at later times ($ > 500-2000$ milliseconds), for it is then that
the emergent luminosities are powered by the energy in the dense core.
At this time, since most of the core pressure is derived from the cold, stiff
nuclear component, and not an ideal gas of non--interacting nucleons,
the effective specific heat is not negative, but positive, and cooling
does not lead to much of a density change.  Therefore, the effect of a decrease
in opacity will not be partially cancelled by a density increase 
and the emergent luminosities will reflect the full suppression effects.
Since the neutrino luminosity is the agency of explosion, this could
be important for a model in which the delay to explosion is not short.
There is an important caveat: accretion of the infalling envelope
is a major power source and its continuance can mask the effects of
the alterations we find in the neutrino opacities at depth.
Accretion will be less important for the lower--mass progenitor stars
($8 - 13\, {\rm M}_{\odot}$),  which have more tenuous envelopes.
In addition, the neutrino signals observed from SN1987A \cite{hir,bion} constrain
the magnitude of any alteration in our standard model, as does the fact
that there is a fixed amount of energy to be radiated from any
given protoneutron star.  In this paper, to gauge these
effects and to begin the debate on these issues, 
we perform an idealized protoneutron star cooling calculation.

In \S II, we explain the physics of the correlation effects
and provide a simple, single--channel, model that makes clear the character 
of the results.  In \S III, the general formulation
is presented, as is an aside on the low--wavelength limit.
In \S IV, the complete correlation and dynamic structure function 
formalism is derived and discussed and a straightforward nucleon
potential model is described.  In reality, the solution of the problem
of the nuclear equation of state goes hand--in--hand with the calculation
of many--body correlations.   Since we are not proposing here to resolve the former,
we are satisfied with a model that captures the essence of the nuclear
interaction.  Different nuclear equations of state will yield 
quantitatively, but not qualitatively, different results.
In \S V, we outline our multi--channel formalism for calculating
$\nu$--nucleon scattering structure functions and in \S VI we present 
the results of our calculations for various temperatures, densities, and neutrino
energies.  In \S II through \S VI, we start from the general expressions of
statistical mechanics that encompass the effects of the medium
on the neutral--current neutrino reactions and proceed to calculations of the effects.
In \S VII, we present the results of a suggestive series
of calculations that might point to the astrophysical import of the
new effects and in \S VIII we summarize our conclusions.

\section{The Essential Physics}

We can introduce, and qualitatively explain, much of what is 
to come in a simplified system in which we have just one species of 
non-relativistic nucleon with mass $m$, which is coupled to neutrinos 
through only a vector neutral current. We take the weak--interaction Lagrangean density to be,

\begin{equation}
{\cal L_W}=\frac{G_W}{\sqrt{2}} \bar{\psi}_{\nu}(x)(1-\gamma_5)\gamma_0\psi_{\nu}(x) n(x),
\label{first}
\end {equation}
where $n(x)$ is the density operator for the nucleons, $G_W$ is the weak coupling constant, 
and the other symbols have their standard field--theoretic meanings.  In the following, we employ
a system of units in which $\hbar=c=k_B=1$.

We write the differential rate of neutrino scattering $\bf p_1\to p_2$ in the medium as

\begin{equation}
\frac{d^2 \Gamma}{d \omega \hspace{1 pt} d \cos\theta}=(4 \pi^2)^{-1}
G_W^2 (E_1-\omega)^2[1-f_{\nu}(E_1-\omega)]\Lambda^{00}(q,\omega) S(q,\omega),
\label{b2}
\end{equation}
where $f_{\nu}$ is the occupation function for neutrinos, $E_1$ is the energy of the 
incident neutrino, $\Lambda$ is the neutrino trace,  

\begin{equation}
\Lambda^{\mu\nu}=(4E_1E_2)^{-1}Tr[p_1 \hspace{-9.5pt}/ \hspace{8 pt}(1-\gamma_5)
\gamma^{\mu}p_2 \hspace{-9.5 pt}/\hspace{8pt}\gamma^{\nu}(1-\gamma_5)], 
\end{equation}
$\omega$ is the energy transfer (${\bf |p_1|-|p_2|}$) to the 
medium, and $q$ is the momentum transfer (${\bf |p_1-p_2|}$), 
related to $\omega$ and $p_1$ through the neutrino
scattering angle, $\theta$, by

\begin{equation}
q=[p_1^2+(p_1-\omega)^2-2p_1(p_1-\omega) \cos\theta]^{1/2}.
\end{equation}

For the case of free nucleons, the nucleon part, $S(q,\omega)$ 
(the {\it dynamic} structure function) is given by

\begin{equation}
S(q,\omega)= 2\int \frac{d^3p}{(2\pi)^3}f(|{\bf  p}| )
(1-f(|{\bf p+q}|))2\pi\delta(\omega+\epsilon_{\bf p}-\epsilon_{\bf p+q}),
\label{b4}
\end{equation}
where $f({\bf |p|})$ is the Fermi--Dirac distribution function and 
$\epsilon_{\bf p}$ is the nucleon energy. Eq.~(\ref{b2}) is  
Fermi's golden rule with final--state nucleon blocking.  The energy 
delta function expresses energy conservation and momentum conservation, 
comes from the matrix elements of the nucleon density operators 
$<p'|\int d^3x\, e^{i \bf q \cdot x \rm\it}\, n(\bf x \rm\it)|p>$,  
and has already been integrated out. From the  
$\delta$ function in (\ref{b4}) we see that a typical energy transfer  
$\omega$ from the medium to the neutrino is of the order of  $q$ times 
the thermal velocity, $(T/m)^{1/2}$.  Thus, in the limit of heavy nucleons, 
when we integrate the differential rate (\ref{b2}) over a range of 
$\omega$s the other factors in the integrand can be evaluated at  
$\omega=0$. We can express this limit as,

\begin{equation}
(2\pi)^{-1}S(q,\omega)\rightarrow
(2\pi)^{-1}\delta (\omega )\int d\omega' S(q,\omega')\equiv \delta (\omega )S(q) ,
\label{b6}
\end{equation}
where $S(q)$ is the {\it static} structure factor.  In \S VI, we show that
at the high densities and temperatures achieved in the supernova context
the $\omega = 0$ (elastic) limit is not particularly accurate (see also \cite{reddy}).

When we turn on interactions among the nucleons, (\ref{b4}) is replaced by,
\begin{equation}
S(q,\omega)=2\pi Z^{-1}\sum_{j,k}e^{-\beta E_j}\int d^3x\, e^{i \bf q\cdot x}
<j|n(x)|k><k|n(0)|j>\delta(\omega+E_j-E_k),
\label{b7}
\end{equation}
where $j$ and $k$ are energy eigenstates, now of the whole medium, 
$Z$ is the partition function, $T$ is the temperature, and $\beta = \frac{1}{T}$.
If the medium is comprised of heavy enough nucleons, 
we need only the static structure function defined in (\ref{b6}), then given by

\begin{equation}
S(q)=Z^{-1}\sum_{j}e^{-\beta E_j}\int d^3x\, e^{i \bf q\cdot x}<j|n({\bf x})n(0)|j>.
\label{2piS}
\end{equation}

In other words, $S(q)$ is merely the Fourier transform of the thermally--averaged
density--density correlation function.  This is the classic result that 
scattering off of a medium is in reality scattering off of the {\it fluctuations}
in that medium.  Also of interest is the long--wavelength limit, $q \rightarrow 0$,
justified when the neutrino wavelength is much bigger than the 
interparticle separation.\footnote{We emphasize that the limits 
discussed in this section are presented both for pedagogical reasons 
and because they provide a boundary condition for the later work 
in this paper, in which we require neither limit. Under the actual 
conditions that prevail in the supernova core the conditions for the 
limits to be applicable are marginally satisfied, at best.} Statistical mechanics
provides two useful and equivalent expressions 
for the long--wavelength limit, $S(0)$, the first \cite{landau}
in terms of the isothermal compressibility of the medium $K_T$ 
($=-\frac{\partial \log V}{\partial P}\bigr|_T$), 

\begin{equation}
S(0)= \bar{n}^2 \beta^{-1} K_T = \bar{n} \frac{K_T}{K_0},
\label{fluct1}
\end{equation}
where $K_0$ is the ideal gas compressibility and 
$\bar{n}$ is the average nucleon density,
and the second ({\it ibid.}, their eq. 114.14) in terms of the derivative of the density with respect 
to the chemical potential of the nucleons, $\mu$,

\begin{equation}
S(0)=  \beta ^{-1}\frac{\partial \bar{n}}{\partial \mu}.
\label{fluct2}
\end{equation}
In the ideal gas limit of no correlation between particles, eqs.~(\ref{fluct1}) 
and (\ref{fluct2}) show that $S(q)$ is simply equal 
to the number density, $\bar{n}$, as expected from 
eq.~(\ref{2piS}) and eq.~(\ref{b4}), without blocking.
Eq.~(\ref{fluct1}) reveals that if $K_T$ is small because the 
matter is stiff, in the long--wavelength limit
the neutrino--matter cross sections are {\it suppressed}.
When we replace eq.~(\ref{first}) by the complete standard model form,
including the axial--vector current and nucleon isospin, we shall
require separate correlation functions for the neutron and the proton,
as well as for spin correlations. These depend upon susceptibilities
that are different from the compressibility, but we shall find
suppression in these terms as well.

Eq.~(\ref{fluct2}), equivalent to eq.~(\ref{fluct1}) by a thermodynamic identity, is 
a powerful result of great generality.
In standard approximation schemes for the many--body problem, the 
distribution function for a nucleon species is given by the 
Fermi--Dirac distribution in which the chemical potential $\mu$ is 
replaced by $\mu-v(\bar{n})$, where $v(\bar{n})$ is the average 
energy of interaction of the nucleon with the other nucleons 
and is a function of the density. Thus, the density is given implicitly by

\begin{equation}
 \bar{n}=2\int \frac{d^3p}{(2\pi)^{3}}[1+e^{\beta[(p^2/(2m)-\mu+v(\bar{n})]}]^{-1}.
\label{a15}
\end{equation}

The expression (\ref{a15}) holds in the Hartree approximation; 
it holds in approaches that introduce mean meson fields instead 
of nuclear potentials; it holds in the Landau Fermi liquid theory (FLT),
subject to the proviso that we use only results in which the derivative 
of the potential $v$ (with respect to the $ \bar{n}$) enters; and 
it holds in approaches using the Skyrme potential. 

Differentiating (\ref{a15}), we can solve for 
$\frac{\partial \bar{n}}{\partial \mu}$ and $S(0)$,

\begin{equation}
S(0)=\beta^{-1}\frac{\partial  \bar{n}}{\partial \mu}=h(\mu)\bigl[1+h(\mu)
\frac{\partial v}{\partial \bar{n}}\bigr]^{-1},
\label{a16}
\end{equation}
where

\begin{equation}
h(\mu)=2 \int \frac{d^3p}{(2\pi)^3}\frac{e^{\beta[p^2/(2m)-\mu+v]}}
{[1+e^{\beta[p^2/(2m)-\mu+v]}]^2}\\
= 2\int \frac{d^3p}{(2\pi)^{3}} f_1(p)(1-f_1(p))
\label{a17}
\end{equation}
and $f_1(p)$ is the Fermi--Dirac function, but 
with the chemical potential displaced by $v$.  If we regard 
particle densities as inputs to our calculations, then the displacement 
of the chemical potential by the nuclear potential is irrelevant, since 
the same difference, $\mu-v$, enters the calculation of the density in 
terms of the chemical potential. Thus, the numerator of (\ref{a16}) 
contains no more than the familiar Pauli blocking effects (for the case $q=0$); 
the denominator contains all of the effect of the interactions.

As an example, consider a two--nucleon potential $V(r)$. In the Hartree approximation, 
the average potential seen by a single nucleon is given by $v=\bar{n}U$, where
$U= \int d^3x V(x)$, and (\ref{a16}) becomes

\begin{equation}
S(0)=h(\mu)\bigl[1+h(\mu)U\bigr]^{-1},
\label{b20}
\end{equation}
the potential providing an enhancement, if negative, and a suppression, if positive.
The latter is the case for high--density nuclear matter.

Extensions of the above considerations to include isotopic spin and 
the axial--vector weak hadronic current were given in reference~\cite{s2Ray}. Closely 
related considerations were given in reference~\cite{iwa} for the case of degenerate 
neutron matter. Both references~\cite{s2Ray} and \cite{iwa} concluded that nuclear 
interactions cause a big decrease in the Gamow--Teller part of the 
neutrino opacity in the regions 
considered in the respective works. However, in addition to being 
fragmentary in the domains that were covered, these works did not address 
two important issues: 1) the errors in doing the integral over $\omega$ by 
assuming that the neutrino parts of the matrix element and phase space 
factors are independent of $\omega$ over the region of dominant 
contribution and,  2) the applicability of the $q=0$ limit. These limits 
are the least justified for the case of trapped electron neutrinos 
in the early dense core, where the momentum transfers in neutrino 
scattering are comparable to the inverse particle spacing.

In the present work we address these deficiencies by using methods 
that incorporate all of the physics of the above discussion and previously 
cited papers, but which calculate the dynamic structure factor as a 
function of ($q,\omega$) and do the integrals over the neutrino variables, 
without further approximation. We also aim to better systematize the 
problem by keeping the connections to textbook many--body formalism 
\footnote{In what follows, we will use a number of formulae from 
Fetter and Walecka\cite{FW}, hereafter referred to as FW.} as clear as possible. 

\section{General Formulation}

To address in complete form the reactions that a neutrino undergoes 
in a medium we need to calculate the appropriate thermal averages of  
the weak--current operators for the particles that comprise the medium. 
We define $j_{\mu}$ as the weak neutral current of the particles in 
the medium, other than the neutrinos, and we assume conditions under 
which all species, except the neutrinos, are in statistical equilibrium. 
The properties of the medium are embodied in the function $W_{\mu\nu}$

\begin{equation}
W_{\mu\nu}(q,\omega)={\cal F}[j_{\mu},j_{\nu}]_{q,\omega}\, ,
\label{a2}
\end{equation}
where we define the correlation functional ${\cal F}_{q,\omega}$ 
of two Heisenberg operators $O_1$ and $O_2$ as 

\begin{equation}
{\cal F}[O_1,O_2]_{q,\omega}=Z^{-1}\int d^4x \;e^{-i \bf q
\cdot \bf x}\rm\it e^{i\omega t}Tr[e^{-\beta( H-\Sigma \mu_iN_i)}O_1(\bf x \rm\it , t)O_2(0,0)],
\label{a3}
\end{equation}
where $Z$ is the partition function.
We can write the differential rate for neutrino 
scattering in terms of $W_{\mu\nu}(q,\omega)$,

\begin{equation}
\frac{d^2 \Gamma}{d \omega \hspace{1 pt} d \cos\theta}=
(4 \pi^2)^{-1}G_W^2 (E_1-\omega)^2[1-f_{\nu}(E_1-\omega)]
\Lambda^{\mu\nu}(q,\omega) W_{\mu\nu}(q,\omega).
\label{a4}
\end{equation}

Note that the leptonic spinor--trace and phase--space factors are
as in the golden rule formula (\ref{b2}).  The distribution functions
for initial and final nucleon states, and the golden--rule delta
function, have all been subsumed in  $W_{\mu\nu}(q,\omega)$.
When the medium is interacting, (\ref{a4}) 
can be taken as the definition of the correct answer, as it is in the 
form that {\it ab initio} thermal field theory gives for neutrino 
transition rates. Alternatively, when a complete set of intermediate 
energy and momentum eigenstates are inserted between the two 
operators in (\ref{a3}), the $d^4x$ integrals give the delta 
functions, the two matrix elements give the square of the T matrix, 
and the thermal factor weights them appropriately, as in (\ref{b7}).

We take the medium to be composed of protons, neutrons, electrons and 
neutrinos and we deal only with the neutrino scattering 
from the nucleons in the medium.\footnote{Of course, neutrino--electron 
scattering through the neutral--current couplings must be taken into 
account in the determination of the total opacity, and for the case 
of electron neutrinos the charged--current interactions with nucleons 
and with electrons must be used as well.}  

For the nuclear contribution, we assume that the nucleons 
remain non-relativistic to write the space components of 
the function $W_{\mu\nu}$ in the form,

\begin{equation}
W_{i,0}=W_{0,i}=0 \hspace{30pt} W_{i,j}(q,\omega) = 
W_A (q,\omega)\delta_{i,j}+W_B(q,\omega)q_i q_j\, .
\end{equation}

We define $W_V=W_{0,0}$.  The lepton trace can be expressed 
in terms of these functions as follows,

\begin{equation}
W_{\mu\nu}\Lambda^{\mu\nu}=(1+\cos \theta)W_{V}+(3-\cos\theta)
W_A+\omega^2(1+\cos \theta) W_B,
\label{a7}
\end{equation}
where, as before, $\theta$ is the scattering angle, $W_{V}$ 
represents the Fermi (vector) term, and 
$W_{A}$ is the Gamow--Teller (axial--vector) term.

The $W_B$ term in eq.~(\ref{a7}) vanishes in the absence of
interactions among the nucleons. In our later parameterization of
the forces, it can become different from zero only when a tensor
force is added to the conventional four central forces. We have
directly evaluated the contribution to the opacity from the tensor
force coming from single neutral--pion exchange and find, for
degenerate matter around nuclear density, $W_B \approx 2 \times
10^{-5} W_A ({\rm MeV})^{-2}$. The tensor terms give an
additional opacity that is a small fraction of that due to the
axial--vector terms ($W_A$), and we drop them in what
follows.

The non--relativistic limits of the standard model neutral current 
to be used in (\ref{a2}) can be expressed in terms of the neutron 
and proton density operators, $n_n$ and $n_p$, and the \^{z} 
component of the spin--density operators,
$n^{(3)}_{n,p}= \psi_{n,p}^{\dag} \sigma_3 \psi_{n,p}$,

\begin{eqnarray}
&j_0(x)=C_V^{(p)} n_p(x)+C_V^{(n)}n_n(x)
\nonumber\\
&j_3(x)=g_A(n_p^{(3)}(x)-n_n^{(3)}(x)),
\end{eqnarray}
where $C_V^{(p)}=1/2-2\sin^2\theta_W$, $C_V^{(n)}=1/2$, 
$g_A = -1.26/2$, and $\sin^2\theta_W = 0.23$.
By rotational invariance, we need only the correlation 
functions of the \^{z} component of the spin, but, since the medium 
is not invariant under isospin rotations, we use separate neutron 
and proton currents.

For the vector (Fermi) part, we now obtain,

\begin{equation}
W_V(q,\omega)=(C_V^{(p)})^2S_{pp}(q,\omega)+
2C_V^{(p)}C_V^{(n)}S_{pn}(q,\omega)+(C_V^{(n)})^2S_{nn}(q,\omega),
\label{a9}
\end{equation}
where the structure functions are defined as $S_{pp}={\cal F}[n_p,n_p]$,
$S_{pn}={\cal F}[n_p,n_n]$, and $S_{nn}={\cal F}[n_n,n_n]$.

For the axial--vector (Gamow--Teller) part, we have

\begin{equation}
W_A(q,\omega)=g^2_A(S^A_{pp}(q,\omega)+S^A_{nn}(q,\omega)-2S^A_{pn}(q,\omega))=g^2_AS_A(q,\omega),
\label{a10}
\end{equation}
where $S^A_{pp}={\cal F}[n^{(3)}_p,n^{(3)}_p]$, 
$S^A_{pn}={\cal F}[n^{(3)}_p,n^{(3)}_n]$ , and $S^A_{nn}={\cal F}[n^{(3)}_n,n^{(3)}_n]$. 
The purpose of the remainder of this paper is to calculate 
the six structure functions. 

\subsection{An Aside on the Long--Wavelength Limit}

In the long--wavelength limit, we can calculate 
the six structure functions quite simply
and intuitively. 
We introduce the average densities, $\bar{n}_i$, 
for the four species, $\bar{n}_i=(Vol)^{-1}\int d^3x\,n_i(x)$, where 
the index $i$ runs over the values $p^{\uparrow}$, $p^{\downarrow}$, 
$n^{\uparrow}$, $n^{\downarrow}$, and we form the combinations,

\begin{eqnarray}
\bar{n}_{p,n}=\bar{n}_{p,n}^{\uparrow}+\bar{n}_{p,n}^{\downarrow}
\nonumber\\
\bar{n}^A_{p,n}=\bar{n}_{p,n}^{\uparrow}-\bar{n}_{p,n}^{\downarrow}.
\end{eqnarray}

We also introduce separate chemical potentials for the up and down 
spin states for the two species and introduce the notations,

\begin{eqnarray}
\mu_{p,n}=\frac{1}{2}(\mu_{p,n}^{\uparrow}+\mu_{p,n}^{\downarrow})
\nonumber\\
\mu^A_{p,n}=\frac{1}{2}(\mu_{p,n}^{\uparrow}-\mu_{p,n}^{\downarrow}).
\end{eqnarray}

Then, as shown in (\ref{fluct2}) and reference~\cite{s2Ray}, 
we can express the $q \rightarrow 0$ 
limits of the static structure functions as follows:

\begin{eqnarray}
&S_{pp}(0)=\beta^{-1}\frac{\partial \bar{n}_p}{\partial \mu_p}
\; \; \;  ;  \; \; S_{nn}(0)=\beta^{-1}\frac{\partial \bar{n}_n}{\partial \mu_n}
\nonumber\\
&S_{pn}(0)=S_{np}(0)=\beta^{-1}\frac{\partial \bar{n}_p}{\partial \mu_n}
\nonumber\\
&S^A_{pp}(0)=\beta^{-1}\frac{\partial \bar{n}^A_p}{\partial \mu^A_p}
\; \; \;  ;  \; \; S^A_{nn}(0)=\beta^{-1}\frac{\partial \bar{n}^A_n}{\partial \mu^A_n}
\nonumber\\
&S^A_{pn}(0)=S^A_{np}(0)=\beta^{-1}\frac{\partial \bar{n}^A_p}{\partial \mu^A_n}.
\label{a14}
\end{eqnarray}

As in the one--channel case treated in \S II,  we take the 
distribution function for a nucleon species, $i$, to be given by the 
Fermi--Dirac distribution in which the chemical potential, $\mu_i$, has been 
replaced by $\mu_i-v_i$, where $v_i$ is the average energy of 
interaction of the nucleon with all other particles. 
Thus, the density is given by

\begin{equation}
 \bar{n}_i=2 \int \frac{d^3p}{(2 \pi)^{3}}[1+e^{\beta(p^2/(2m)-\mu_i+v_i)}]^{-1}.
\label{a15prime}
\end{equation}

In general, the potential, $v_i$  for species $i$ is a function 
of the densities for all four of the species, $ \bar{n}_j$. If we know 
the functional dependence of the $ v$'s on the $ \bar{n}_j$s, we can solve 
for the long--wavelength limit of the Fermi static structure functions, $S$, 
by differentiating the four equations (\ref{a15prime}), with respect to 
the four chemical potentials that we have introduced. We give the 
solution for the two combinations of structure functions that enter our rate calculations.
In (\ref{a9}), with $\sin^2\theta_W=1/4$, we have only the neutron contribution,

\begin{equation}
S_{nn}=\beta^{-1}\frac{\partial \bar{n}_n}{\partial\mu_n}=h(\mu_n)/d_V,
\label{sv43}
\end{equation}
where

\begin{equation}
d_V=1+\frac{\partial v_n}{\partial \bar{n}_n}h(\bar{n}_n)
-\Bigl[ \frac{\partial v_n}{\partial \bar{n}_p}\Bigr]^2 h(\mu_n)h(\mu_p)
 \Bigl[1+\frac{\partial v_p}{\partial \bar{n}_p}h(\mu_p)\Bigr]^{-1}.
\end{equation}

In (\ref{a10}), integrated over $\omega$, we obtain a simple form, 
if we choose potentials $v_i$ such that $\partial v_i/\partial\bar{n_j}=
\partial v_j/\partial\bar{n_i}$ and such that $v_{p\uparrow}-v_{p\downarrow}
+v_{n\uparrow}-v_{n\downarrow}=0$, two conditions that are fulfilled 
by the potentials we use. We then obtain,

\begin{equation}
S_A(q)=S^A_{pp}(q)+S^A_{nn}(q)-2S^A_{pn}(q)=\frac{h(\mu_p)+h(\mu_n)}{1+v_a(h(\mu_p)+h(\mu_n))},
\label{sa45}
\end{equation}
where $v_a=\frac{1}{8}\bigl[\frac{\partial}{\partial \bar{n}^A_p}-\frac{\partial}
{\partial \bar{n}^A_n}\bigr]\bigl[v_{p\uparrow}-v_{p\downarrow}-v_{n\uparrow}+v_{n\downarrow}\bigr]$.

We present these static, long--wavelength results~(eqs.~\ref{sv43} and \ref{sa45}),
which depend on effective single--particle potentials, because they show a
strong similarity in form to the more complete equations that will be
developed in the next section and because, in most regions, they give numerical results
that are not greatly different. However, it should be pointed out that any
program that begins with potentials that fit nuclear data and calculates
the ground--state properties of nuclear matter ({\it cf.}, \cite{fp,sp}) is capable, if subjected to
the right sets of constraints, of directly determining the low--temperature
values of the static structure functions enumerated in (\ref{a14}), without recourse to
the assumptions that led to eqs.~(\ref{sv43}) and (\ref{sa45}). Implementation requires use of
the multichannel analogues of the connection between the one--species
structure function and the bulk modulus (\ref{fluct1}). The required numerical
experiment involves constraining the system to have different expectation
values of proton spin and neutron spin densities (in the $\hat z$
direction), calculating the constrained ground--state energy as a function
of these densities (as well as the particle densities), and taking
combinations of second derivatives of this energy with respect to the
densities, evaluated finally for the values of the densities in the true
ground state. We strongly recommend that groups that do nuclear matter
calculations carry out these steps.

\section{Complete Correlation Functions}

In the field--theoretic formalism for the quantum mechanical many--body 
problem there are well--formulated perturbative techniques for directly 
calculating correlation functions, without restricting to the equal 
time ($\omega = 0$) and $q \rightarrow 0$ limits. In Appendix B, we show that the 
limiting results sketched out in the previous section by calculating 
the derivatives, $\frac{\partial \bar{n_i}}{\partial \mu_i}$, are exactly what 
we obtain, in the same limits, from the sum of the ``ring" 
approximation graphs in this many-body formalism. Therefore, we must  
sum at least the ring graphs to recapture the long--wavelength limit 
of the static structure function. In so doing, we obtain the full 
dynamic structure function $S(q,\omega)$ at the same level at which we find  
the limiting forms in \S III.A.

To follow a graph summing approach we must replace the 
``correlation functional" of two (bosonic) operators defined 
in (\ref{a3}), $ \cal F$, by a retarded commutator form, $ \tilde{\cal F}$,

\begin{equation}
\tilde{\cal F}[O_1,O_2]_{q,\omega}=-\frac{i}{Z}\int d^4x \;e^{-i \bf q
\cdot \bf x}\rm\it e^{i\omega t}Tr\bigl[e^{-\beta( H-\Sigma 
\mu_i N_i)}[O_1(\bf x \rm\it , t),O_2(0,0)]\bigr]\theta (t),
\label{a13}
\end{equation}
where here $\theta(t)$ is the Theta function.

The ``correlation functional," eq.~(\ref{a3}), is 
recaptured through (FW eqs. 32.14 and 32.16),

\begin{equation}
{\cal F}_{q,\omega}=2 {\rm Im}[\tilde{\cal F}_{q,\omega}](1-e^{-\beta\omega})^{-1}.
\end{equation}

Reverting to the single--channel problem for illustrative 
purposes, we define the polarization function as

\begin{equation}
\Pi(q,\omega)=\tilde{\cal F}[n,n]_{q,\omega}\, ,
\label{a22}
\end{equation}
so that 

\begin{equation}
S(q,\omega)=2 {\rm Im} [\Pi(q,\omega)](1-e^{-\beta\omega})^{-1}.
\label{a21}
\end{equation}

Note that we can use the fact that $ {\rm Im} \Pi(q,\omega)$ is odd 
in the variable, $\omega$, to derive from (\ref{a21}) the 
relation that embodies detailed balance, $S(-\omega)=e^{-\beta\omega}S(\omega)$.
The ring approximation (sometimes referred to as the 
Random Phase Approximation, RPA) then gives

\begin{equation}
\Pi(q,\omega)=\frac{\Pi^{(0)}(q,\omega)}{1-v(q) \Pi^{(0)}(q,\omega)},
\label{a23}
\end{equation}
where $v(q) =\int d^3 x V(x)$ and $\Pi^{(0)}(q,\omega)$ is the free polarization.
Eq.~(\ref{a23}) is of the same
general form as eq.~(\ref{a16}) for the equal--time and small--$q$
limits. If we take only the numerator in (\ref{a23}),
we recover the effects of Pauli blocking alone. 

To complete the calculation, we need the real and imaginary parts 
of $\Pi^{(0)}(q,\omega)$, which we derive in Appendix A:

\begin{equation}
{\rm Re} \;\Pi^{(0)}(q,\omega)=\frac{m^2}{2\pi^2q \beta}\int_0^{\infty} 
\frac{ds}{s}\log  \Bigl[\frac{1+e^{-(s+Q)^2+\beta \mu}}{1+e^{-(s-Q)^2+\beta \mu}} \Bigr]\; \; 
+(\omega \rightarrow -\omega)
\label{a24}
\end{equation}

and

\begin{equation}
{\rm Im} \Pi^{(0)}(q,\omega)=\frac{m^2}{2\pi \beta q} \log 
\Bigl[\frac{1+e^{-Q^2+\beta \mu}}{1+e^{-Q^2+\beta \mu-\beta\omega}} \Bigr],
\label{a25}
\end{equation}
where

\begin{equation}
Q=\bigl(\frac{m\beta}{2}\bigr)^{1/2} \bigl(-\frac{\omega}{q}+\frac{q}{2m} \bigr)
\end{equation}
and for $m$ we should substitute the effective mass, $m^*$\cite{reddy}.

To obtain the various $S(q,\omega)$ functions that are used 
in (\ref{a23}) for determining the rates, the above formulae are 
generalized to the multiple--channel case by considering the 
correlations of every pair of densities from the set $n_p$, 
$n_n$, $n^3_p$,$n^3_n$. The cross correlations 
between the densities and the spin densities, $n^3_{i}$, vanish, as do 
all dynamical connections between the two at the ring level, so 
that we are left with two $2\times2$ problems in solving for the correlations. 
In each sector, we define a matrix polarization function 
$\Pi(q,\omega)_{i,j}=\tilde{\cal F}[n_i,n_j]_{q,\omega}$ and a 
ring approximation, which is the simple matrix extension of (\ref{a23}). 
For the free polarization matrix, $\Pi^{(0)}$, we take a diagonal 
matrix in which the proton and neutron elements are given 
by (\ref{a24}) and (\ref{a25}), with the respective proton and 
neutron chemical potentials,  $\mu_p$ and $\mu_n$, in place of $\mu$. 
The generalization of the potential, $v$, will connect the proton 
and neutron elements. The elements of the structure function matrices,  
$S$ and $S^A$, used in the rate formulae follow immediately 
from the matrix form of (\ref{a23}). It remains to address the potentials.

\subsection{Nucleon Potential Model}

A general velocity--independent local two--body interaction 
between nucleons (a) and (b) is given by, 

\begin{equation}
V^{a,b}=V_1(r)+ \bf\vec{\tau} \it^{ a} \cdot\vec{\tau}^b  V_2(r)+\vec{\sigma}^a 
\cdot\vec{\sigma}^b \it V_3(r)+\vec{\tau}^a 
\cdot\vec{\tau}^b\vec{\sigma}^a \cdot\vec{\sigma}^b V_4(r).
\label{a30}
\end{equation}

This leads directly to the following construction: 
\footnote{In interpretating these matrix elements it should be borne 
in mind that they are not analogous to matrix elements of an operator 
between single--particle states. They operate in a space of two neutral 
densities (the arguments of $\tilde{\cal F}$) and describe the 
scattering of two particles, with no charge exchange.}

a) to determine the Fermi $S(q,\omega)$ elements, we use 

\begin{eqnarray}
&v_{pp}(q)=v_{nn}(q)=v_1(q)+v_2(q)
\nonumber\\
&v_{pn}(q)=v_{np}(q)=v_1(q)-v_2(q)
\label{a28}
\end{eqnarray}

and
\vspace{12pt}
b) to determine the Gamow--Teller $S^A(q,\omega)$ elements, we use

\begin{eqnarray}
&v^A_{pp}(q)=v^A_{nn}(q)=v_3(q)+v_4(q)
\nonumber\\
&v^A_{pn}(q)=v^A_{np}(q)=v_3(q)-v_4(q).
\label{28}
\end{eqnarray}

Of course, in the nuclear interaction problem within nuclei, 
as well as at the higher densities in neutron star matter, there are 
a few complications: 1) there isn't a potential,  2) taking one anyway, 
it is too strong to allow the use of perturbation theory \footnote{For example, 
our plane--wave Hartree approximation, used with the potential that is 
attributed to $\omega$--meson exchange, would give much too much 
positive energy and too great a rate reduction for the Fermi terms, since it 
doesn't keep the particles apart at short distances.}, 3) the better 
numerical methods ({\it e.g.}, correlated basis functions)  
for determining  the ground--state energy of nuclear matter from 
a phenomenological potential may be poorly adapted to 
calculation of the dynamic structure functions.

Since it is the excitations of the medium, rather than the cold 
equation of state, that really enter this problem, the Landau Fermi 
liquid theory provides a framework for proceeding. This 
theory defines an energy functional associated with variations of the 
various densities. We define combinations of density variations,

\begin{eqnarray} 
&\delta_{0,0}=\delta n_p+\delta n_n 
\nonumber\\
&\delta_{3,0}=\delta n_p-\delta n_n 
\nonumber\\
&\delta_{0,3}=\delta n_p^{\uparrow}+\delta n_n^{\uparrow}-\delta n_p^{\downarrow}-
\delta n_n^{\downarrow}  
\nonumber\\
&\delta_{3,3}= \delta n_p^{\uparrow}-
\delta n_n^{\uparrow}-\delta n_p^{\downarrow}+
\delta n_n^{\downarrow}, 
\end{eqnarray}
where the vertical arrows signify spin--up and spin--down densities. 
The absence of an arrow implies the sum of these densities.

The energy response to these variations that defines the FLT is given by

\begin{eqnarray}
\delta E = \sum_{q,i}\epsilon_i^{(0)}(q)\delta  n_i (q)+ 
 \frac{\lambda}{2}\sum_{q,q'}\bigl[\delta_{0,0}({\bf q}) \delta_{0,0}({\bf q'})F+
\nonumber\\
\delta_{3,0}({\bf q})\delta_{3,0}({\bf q'})F'
+\delta_{0,3}({\bf q})\delta_{0,3}({\bf q'})G+\delta_{3,3}({\bf q})\delta_{3,3}({\bf q'})G'\bigr],
\label{lamb}
\end{eqnarray}
where $\lambda=\pi^2(2m^*p_F)^{-1}$ and $p_F$ is the nucleon Fermi momentum.
The FLT parameters are usually a function of the angle between $q$ and $q'$. 
We shall consider the wavelength in our application to be long enough 
so that only the S--wave parameters, $F_0,F_0',G_0,G_0'$, enter. 
Then, we note that the interaction term, quadratic in the $\delta$s, 
is exactly what one would get from the form (\ref{a30}), with  
zero--range two--body potentials, where we define $v_i=\int d^3r V_i(r)$, and obtain

\begin{equation}
v_1=\lambda F_0; \quad v_2=\lambda F_0'; \quad v_3=\lambda G_0; \quad v_4=\lambda G_0'\, .
\label{a34}
\end{equation}
Note that all the $v_i$s are real.

The parameter $\lambda$ in (\ref{lamb}) implies 
that our potential has a density dependence. However, we view this as
simply part of the mechanics of fixing our parameters $\int d^3rV_i$ 
in terms of the hodgepodge of nuclear phenomenology, at nuclear densities, 
and meson--exchange considerations that went into the fixing of the FLT 
parameters. We keep the parameter $p_F$ fixed at its value for nuclear 
density, and use our derived parameters at all densities. This is 
clearly a sounder procedure at or near nuclear densities, where 
unextrapolated phenomenology was the main input, but we shall calculate 
the results over a wide region of densities nonetheless. We note that 
the use of Skyrme interactions, expressing an effective energy functional 
as a form quadratic and cubic in the densities, would be subject to the same caveats.

For supernova and protoneutron star applications, we must 
include Coulomb forces in our interaction Hamiltonian. 
We can do so by adding to the (zero--range) proton-proton force 
in (\ref{a28}), a Thomas--Fermi screened Coulomb force between protons,

\begin{equation}
v_{pp}=\int d^3r V^{p,p}=v_1+v_2 \rightarrow v_1+v_2+4 \pi e^2 (q^ 2+q_{TF}^2)^{-1},
\label{a33}
\end{equation}
where $q_{TF}^2=4 e^2\pi^{1/3}(3\bar{n}_p)^{2/3}$.
In the denser regions of the star, the screening momentum 
is larger than the $q$ for typical neutrino scattering. In this case, 
the Coulomb term in (\ref{a33}) for the proton-proton interaction 
is independent of $q$ and $e^2$. \footnote{The physics of this additional 
term can best be elucidated by considering for a moment what the effects 
would be if the Fermi term in our weak current coupled only to the protons 
in the medium (rather than almost entirely to the neutrons, as in 
the standard model.) In this case the scattering of neutrinos would be 
dominated by the scattering from proton--density fluctuations on the 
order of the neutrino wavelength. For long neutrino wavelengths, such 
fluctuations are strongly suppressed by the Coulomb force. When $q<<q_{TF}$ 
the price paid in energy for the fluctuation is measured by the increase 
in the kinetic energies of the neutralizing electrons, and we note that 
in this limit the Coulomb term in (\ref{a33}) is just the second 
derivative of the electron Fermi energy with respect to electron 
(or proton) density. In the case at hand, where $\sin\theta_W\approx 1/2$ 
so that the weak coupling is entirely with the neutrons, the Coulomb 
force between protons turns out nonetheless to be quite important, 
because of the strong coupling of neutrons to protons in the symmetry--energy term.} 

\section{Final Procedure for Calculating $\nu$--nucleon Structure Functions}

From eqs.~(\ref{a4}), (\ref{a7}), (\ref{a9}), and (\ref{a10}) and taking 
only the neutron part of the vector--current coupling, the 
differential scattering rate is given by,

\begin{eqnarray}
\frac{d^2 \Gamma}{d \omega \hspace{1 pt} d \cos\theta}=
(4 \pi^2)^{-1}G_W^2 E_2^2[1-f_{\nu}(E_2)]\Bigl[\bigl(1+
\cos\theta\bigr)(C_{V}^n)^2S_{nn}(q,\omega)
\nonumber\\
+\bigl(3-\cos\theta\bigr)g_{A}^2\bigl[ S^A_{pp}(q,\omega)+
S^A_{nn}(q,\omega)-2S^A_{pn}(q,\omega)  \bigr],
\label{a38}
\end{eqnarray}
where $E_2$=$E_1-\omega$.

The structure functions, $S$ and $S^A$, are elements of separate 
$2\times2$ symmetric matrices. For the vector dynamic structure 
function, $S$, we have

\begin{displaymath}
S(q,\omega)=
\pmatrix{S_{pp}(q,\omega)&S_{pn}(q,\omega)\cr S_{pn}(q,\omega)&S_{nn}(q,\omega)\cr}.
\end{displaymath}

The structure function matrix is given by,

\begin{equation}
S(q,\omega)=2 {\rm Im} \Bigl[\Pi^{(0)}(q,\omega)
[1-v(q) \Pi^{(0)}(q,\omega)]^{-1}
\Bigr](1-e^{-\beta\omega})^{-1}
\label{ebetao}
\end{equation}
where

\begin{displaymath}
\Pi^{(0)}(q,\omega)=
\pmatrix{\Pi^{(0)}_p(q,\omega) &0\cr 0&\Pi^{(0)}_n (q,\omega)\cr}
\end{displaymath}
and $\Pi^{(0)}_p$ and $\Pi^{(0)}_n$ are given by the polarization function, 
defined in (\ref{a24})and (\ref{a25}) and evaluated with 
the proton and neutron chemical potentials, respectively.
The potential matrix is,

\begin{displaymath}
v=\pmatrix{v_1+v_2+4\pi e^2 (q^2+q_{TF}^2)^{-1}&v_1-v_2\cr v_1-v_2&v_1+v_2\cr}, 
\end{displaymath}
where the $v$'s were defined in terms of Fermi liquid parameters 
in (\ref{a34}).

In a real calculation, in all the kinematic expressions $m$ is to be
replaced by $m^*$. Unfortunately, the relation of Landau parameters to
experimental results depends upon the effective mass in model--dependent
ways. Taking  $m^*=0.75 m_n$ as our fiducial value for the effective mass,
we use parameters from references~\cite{s4Ray,s5Ray}: $F_0= -0.28; F_0'=0.95; G_0=0;
G_0'=1.7$ and $\lambda=2.63 \times10^{-5} {\rm MeV}^{-2}$, obtaining,

\begin{eqnarray}
&v_1=-7.4 \times 10^{-6}\,{\rm MeV}^{-2}
\nonumber\\
&v_2=2.5\times 10^{-5}\,{\rm MeV}^{-2}
\nonumber\\
&v_3=0
\nonumber\\
&v_4=4.5\times 10^{-5}\,{\rm MeV}^{-2}.
\label{asabove}
\end{eqnarray}

For other values of the effective mass, we keep these potentials at the
same value, which is to say we assume that the Landau parameters are
proportional to  $m^*/ m$. For the dominant spin--independent term,
$v_2$, this accords with the conventional wisdom that the symmetry
energy per nucleon can be written in the form $\alpha((A-Z)/N)^2$. 

The form for the Gamow--Teller matrix, $S^A(q,\omega)$, is the same as that for $S$, 
except that the potential matrix is replaced by $v^A$

\begin{displaymath}
v^A=\pmatrix{v_3+v_4&v_3-v_4\cr v_3-v_4&v_3+v_4\cr}.
\end{displaymath}

Taking the matrix inverses leads to the following forms for the 
combinations of structure functions that appear in (\ref{a38})

\begin{equation}
S_{nn}(q,\omega)=2 {\rm Im} \bigl[\Pi_n^{(0)}D_V^{-1}\bigl](1-e^{-\beta\omega})^{-1},
\label{vector}
\end{equation}
where

\begin{equation}
D_V=1-(v_1+v_ 2)\Pi_n^{(0)}-(v_1-v_2)^2\Pi_n^{(0)}\Pi_p^{(0)}
[1-4\pi e^2(q^2+q_{TF}^2)^{-1}\Pi_p^{(0)}-(v_1+v_2)\Pi_p^{(0)}]^{-1},
\label{DVterm}
\end{equation}
which corresponds in the $(q,\omega) \rightarrow (0,0)$ limits to eq.~(\ref{sv43}).

If, as in (\ref{asabove}), we take $v_3 =0$, we obtain 
the simple result for the axial--current terms,

\begin{equation}
S_A(q,\omega)=
2 {\rm Im} \Big[\frac{\Pi_p^{(0)}(q,\omega)+\Pi_n^{(0)}(q,\omega)}
{1-v_4[\Pi_p^{(0)}(q,\omega)+\Pi_n^{(0)}(q,\omega)]}\Bigr](1-e^{-\beta\omega})^{-1},
\label{axial}
\end{equation}
which corresponds in the $(q,\omega) \rightarrow (0,0)$ limits to eq.~(\ref{sa45}).

\section{Results}

The formulae that we have developed to calculate dynamic structure functions
and scattering rates ({\it i.e.}, cross sections)
for neutrino--matter scattering in nuclear matter, 
in particular eqs.~(\ref{a38}), (\ref{vector}),
and (\ref{axial}), can now be used to obtain quantitative results.
One is free to insert whatever parameters for whatever thermodynamic states
and nuclear models, but we here choose to focus 
on a generic subset of possibilities 
to demonstrate the character of the new results.   
Since there are six neutrino
species in thermal equilibrium in the supernova 
core and the electron types have very different
chemical potentials than the $\nu_\mu $s and 
$\nu_\tau $s, we drop the blocking term,
$ (1-f_{\nu}(E_2)) $, in calculating the total suppression 
factors.   This term is trivial to include in the
general case, and for the $\nu_\mu $s and $\nu_\tau $s 
its omission introduces only a small error. 
However, we want to avoid expanding the number 
of comparisons unduly and the reader is
free to employ the derived equations to calculate 
everything for any combination of parameters.
Since for the differential cross sections one 
does not integrate over the energy transfer,
$\omega$,  and we present not absolute cross 
sections, but cross sections normalized
to the non--interacting case, the differential 
cross section results are fully general. 
Final--state nucleon blocking is always included.

For the Fermi term, since $ C_V^{(p)}=1/2-2\sin^2\theta_W\sim 0 $, 
we drop the proton structure function in
(\ref{a38}).  Furthermore, we use the potential parameters 
given in eq.~(\ref{asabove}), and in eq.~(\ref{DVterm})
we drop the third term.  This term would have 
been significant had it not been for the
Coulomb term in the denominator, an illustration of the importance
of the explicit inclusion of Coulomb forces, even for the neutron
density correlations.   
Since the $v_i$s are all real,
we obtain for the structure factors used in (\ref{a38}),

\begin{equation}
S_{F}(q,\omega)=2 {\rm Im} \Pi_n^{(0)}(1-e^{-\beta\omega})^{-1}{\cal C_V}^{-1},
\label{vectel}
\end{equation}
where
\begin{equation}
{\cal C_V}= (1 - v_F{\rm Re}\Pi_n^{(0)})^2 + 
v_F^{2}({\rm Im}\Pi_n^{(0)})^2,
\label{vectel2}
\end{equation}
and

\begin{equation}
S_A(q,\omega)=
2 \Bigl[{\rm Im}\Pi_p^{(0)}(q,\omega)+{\rm Im}\Pi_n^{(0)}(q,\omega)\Bigr]
(1-e^{-\beta\omega})^{-1}{\cal C_A}^{-1},
\label{axel}
\end{equation}
where
\begin{equation}
{\cal C_A}=
\Bigl[1-v_{GT}({\rm Re}\Pi_p^{(0)}(q,\omega)+{\rm Re}\Pi_n^{(0)}(q,\omega))\Bigr]^2+
v_{GT}^{2}\Bigl[{\rm Im}\Pi_p^{(0)}(q,\omega)+{\rm Im}\Pi_n^{(0)}(q,\omega)\Bigr]^2 \, .
\label{axel2}
\end{equation}

The $F$ in $ S_{F}(q,\omega) $ and the $A$ in $ S_A(q,\omega) $ stand for Fermi
and Gamow--Teller (axial) and $v_F$ and $v_{GT}$ equal 
$(v_1+v_ 2)$ and $v_4$, respectively.  ${\cal C_{V,A}}$ is the correction factor due to 
many--body effects for a given momentum transfer (or scattering angle) and energy 
transfer.   

\subsection{Collective Excitations of the Medium}

\begin{figure}
\vspace*{-1in}
\begin{center}
\leavevmode
\epsfysize=1.00\hsize
\epsfbox[160     160   612   792]{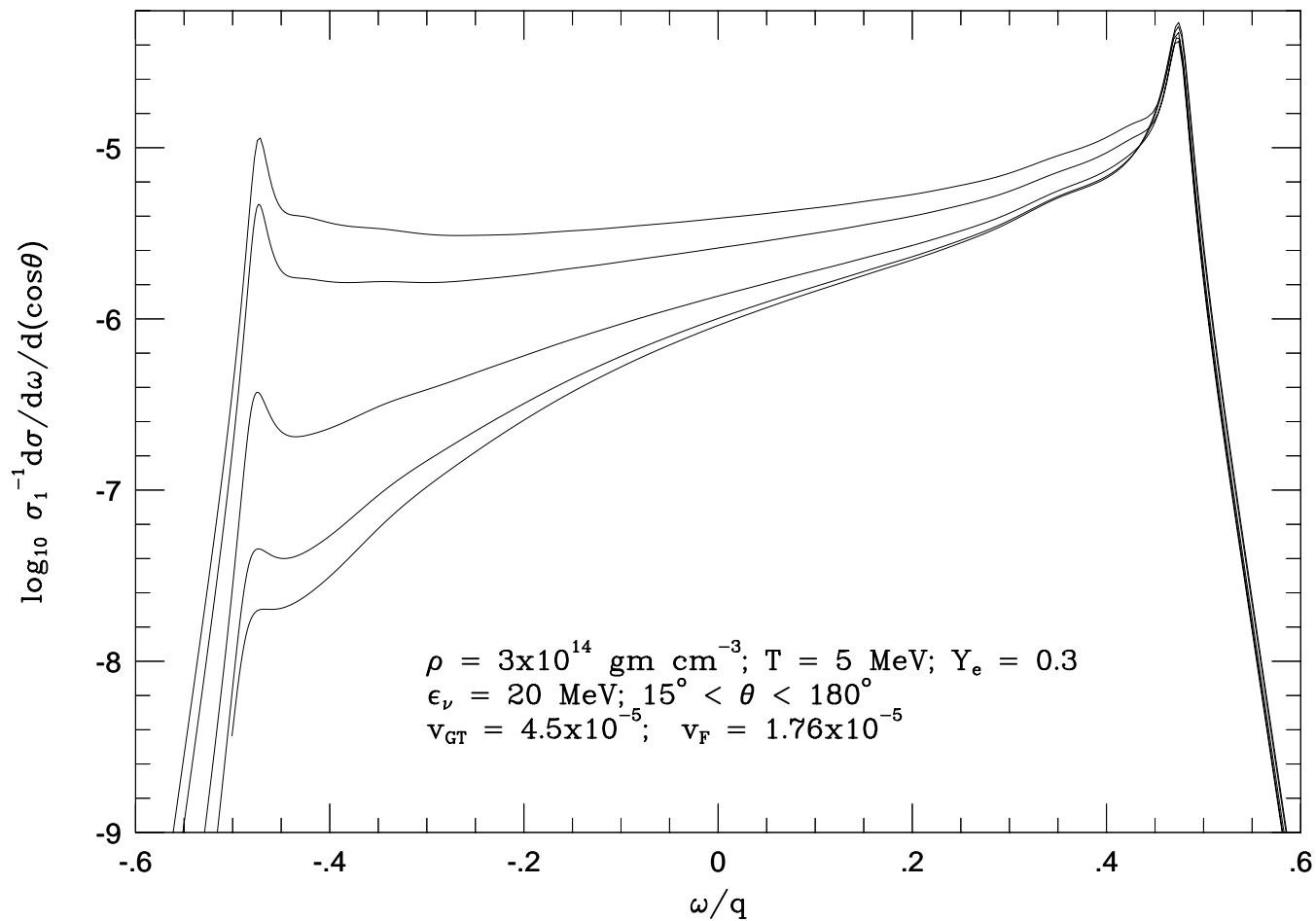}
\end{center}
\vspace*{1in}
\caption{Log$_{10}$ of the doubly--differential cross section 
for neutral--current neutrino--nucleon
scattering versus $\omega/q$ for scattering angles 15$^\circ$, 
45$^\circ $, 90$^\circ $, 135$^\circ $, and 180$^\circ $ .
The calculations were performed at a temperature of 5 MeV, a $Y_e$ of 0.3, a $\rho$
of $3\times10^{14}$ gm cm$^{-3}$, and an incident neutrino energy of 20 MeV.
The default potentials  ($v_{GT}=4.5\times10^{-5}$ 
and $v_F=1.76\times10^{-5}$) 
and effective mass ($m^*=0.75\, m_n$) were employed. 
The differential cross section is divided by the total
scattering cross section ($\sigma_1$) in the non--interacting, 
no--nucleon--blocking, 
$\omega=0$ limit. 
}
\label{fig1}
\end{figure}

\begin{figure}
\begin{center}
\leavevmode
\epsfysize=1.00\hsize
\epsfbox[160     160   612   792]{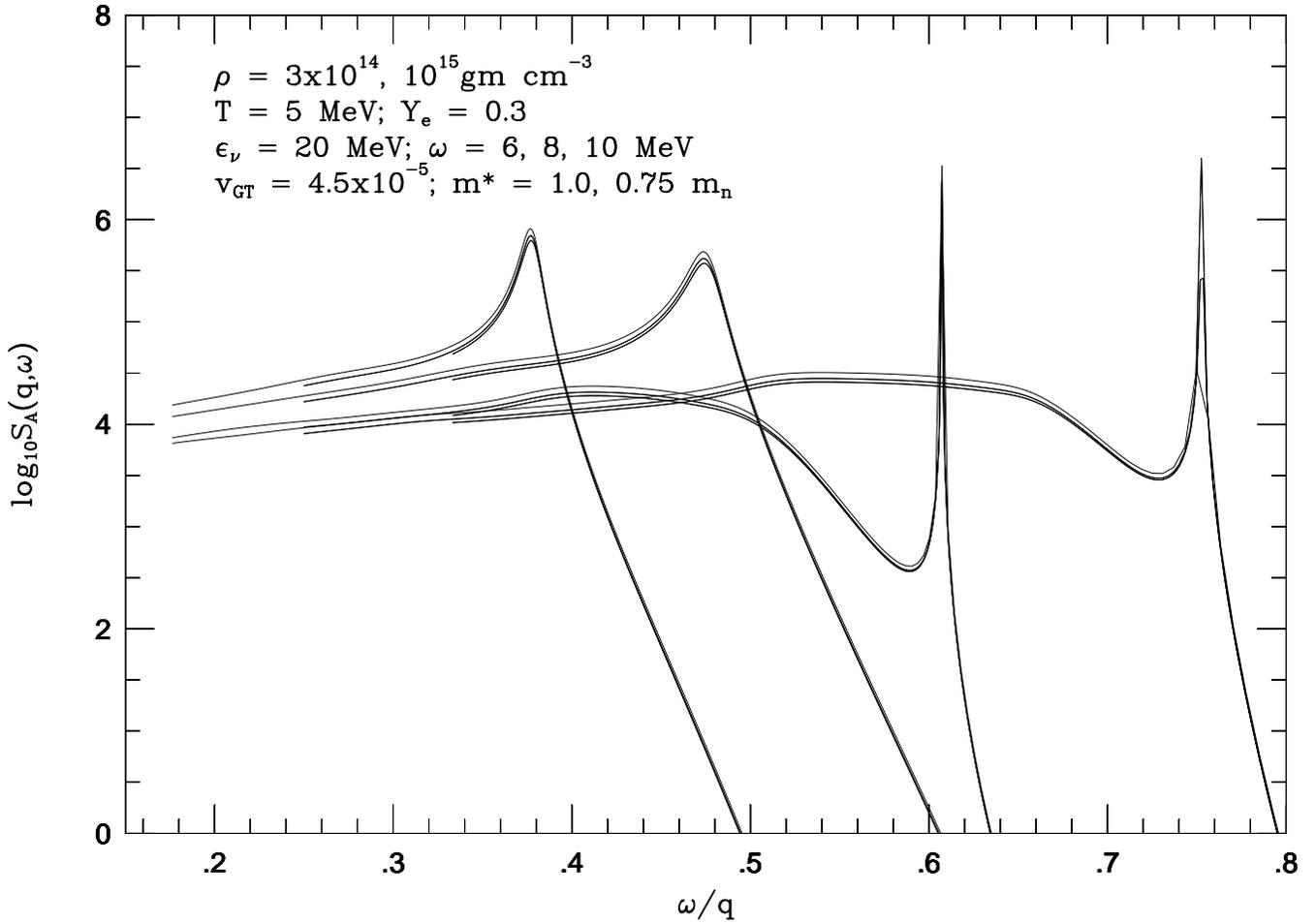}
\end{center}
\vspace*{1.5in}
\caption{Log$_{10}$ of the Gamow--Teller structure function 
versus $\omega/q$ for an incident neutrino energy
of 20 MeV, energy transfers, $\omega$, of 6, 8, 
and 10 MeV, two values of the effective
mass ($ m^* = [0.75m_n, 1.0 m_n]$) and two values 
of the density ($\rho = 3\times10^{14}$ and $10^{15}$ gm cm$^{-3}$).
A temperature of 5 MeV and a $Y_e$ of 0.3 were used, 
as was the default $v_{GT}$ ($=4.5\times10^{-5}$).
}
\label{fig2}
\end{figure}

For most regions of phase space, ${\cal C_A}$ and ${\cal C_V}$ are greater 
than one and represent suppression in the
scattering rates.  Their effects on the integrals 
over $\omega$ and $\theta$ are always
suppressive.  However, the terms containing 
the real parts have roots; these roots represent
collective excitations.  For the Fermi term, zero 
sound in the medium can be generated if the 
scattering has a ($\omega,q$) pair that satisfies 
the mode's dispersion relation, {\it i.e.}, if
it hits the resonance. Similarly, for the Gamow--Teller 
term, spin waves in the protons and the neutrons
(related by a set phase) can be generated.   These 
modes are the traveling--mode equivalents of the
Gamow--Teller resonance in nuclei (a standing wave).  
The zero sound of the Fermi part 
is analogous to the Giant--Dipole resonance in nuclei.  
The resonances increase the structure function when
the scattering transfer ratio, $\omega/q$, equals 
the ratio of the collective excitation's angular frequency
and wavenumber.  For a given scattering angle, one 
can plot the differential cross section in $\omega$
and $\cos\theta$ as a function of $\omega/q$ to see 
the resonances.  In Figure~(\ref{fig1}), we display this
for five different angles between 15$^\circ $ and 180$^\circ $, an 
incident neutrino energy of 20 MeV, a temperature
of 5 MeV, a density of $3\times10^{14}$ gm cm$^{-3}$, 
and an electron fraction, $Y_e$, of 0.3.
We see in Figure~\ref{fig1} that the resonances in 
both the forward and the backward directions line
up at the same values of $\omega/q$, as expected 
for a collective mode, and we can straightforwardly 
calculate the mode's dispersion relation.  
This is akin to the \v{C}erenkov effect.  
Note that the Gamow--Teller term dominates the Fermi term,
so that in Figure~\ref{fig1} we are really seeing the 
spin waves related to the Gamow--Teller resonance.  However, the 
dispersion relations for zero sound and these 
spin waves are generally similar.  In fact,
recalling the classic result \cite{FW} that in the weak--coupling limit, 
the speed of zero sound in a degenerate system is $\sim v_{fermi}$, where
$v_{fermi}$ is the Fermi velocity, and recalling that for nucleons 
in nuclei $v_{fermi}$ is $\sim0.3c$, the calculated 
resonance value of $\omega/q$ is not unexpected.
In Figure~\ref{fig2}, we plot the Gamow--Teller 
structure function versus $\omega/q$ for various
values of $\omega$, $m^*$, and two values of the density.  At $m^*=m_n$, 
for each value of the density we obtain a sharp 
resonance, but at two different speeds,
reflecting the crude $\rho^{1/3}$--dependence 
expected for $v_{fermi}$.  For a given density,
the mode speed is seen in Figure~\ref{fig2} to be 
inversely proportional to the effective mass.
The width of the resonance is determined by the magnitude 
of the imaginary part of the polarization function.

\subsection{Differential Scattering Cross Sections and Suppression Factors}

To calculate the singly--differential scattering cross 
sections (${d\sigma}/{d\omega}$) , we must integrate
(\ref{a38}) over $\cos\theta$.  Since for a given $\omega$ and incident
neutrino energy, $E_1$, this integration is also over a range of $q$s, 
in the process we are smoothing over resonances.
As a consequence, there is no obvious direct signature of 
them in the final result.  It is the doubly--differential
structure functions and cross sections that retain 
the character of the collective modes and resonance.
The integral over the singly--differential scattering cross section 
yields the total cross section, and this can be compared to that
without correlations to gauge the magnitude of the suppression
of the rate.  To demonstrate the nature of the correlation effects,
we have opted to present figures and tables for a subset of the 
possible $\rho$-$T$-$Y_e$-$E_1$ combinations.  Recall that for this
purpose we have employed the default potentials 
(\ref{asabove}) and effective mass ($m^*=0.75 m_n$).
Note, however, that the formalism we have derived is fully general, 
as long as the nucleons can be assumed to be non--relativistic
and we are in the perturbative limit. 
Since we calculate at given neutron and proton densities,
$\mu_n$ and $\mu_p$ are implicit in the formalism, 
as would be the shifts in them due to interactions \cite{reddy}.  

\begin{figure}
\begin{center}
\leavevmode
\epsfysize=1.00\hsize
\epsfbox[160     160   612   792]{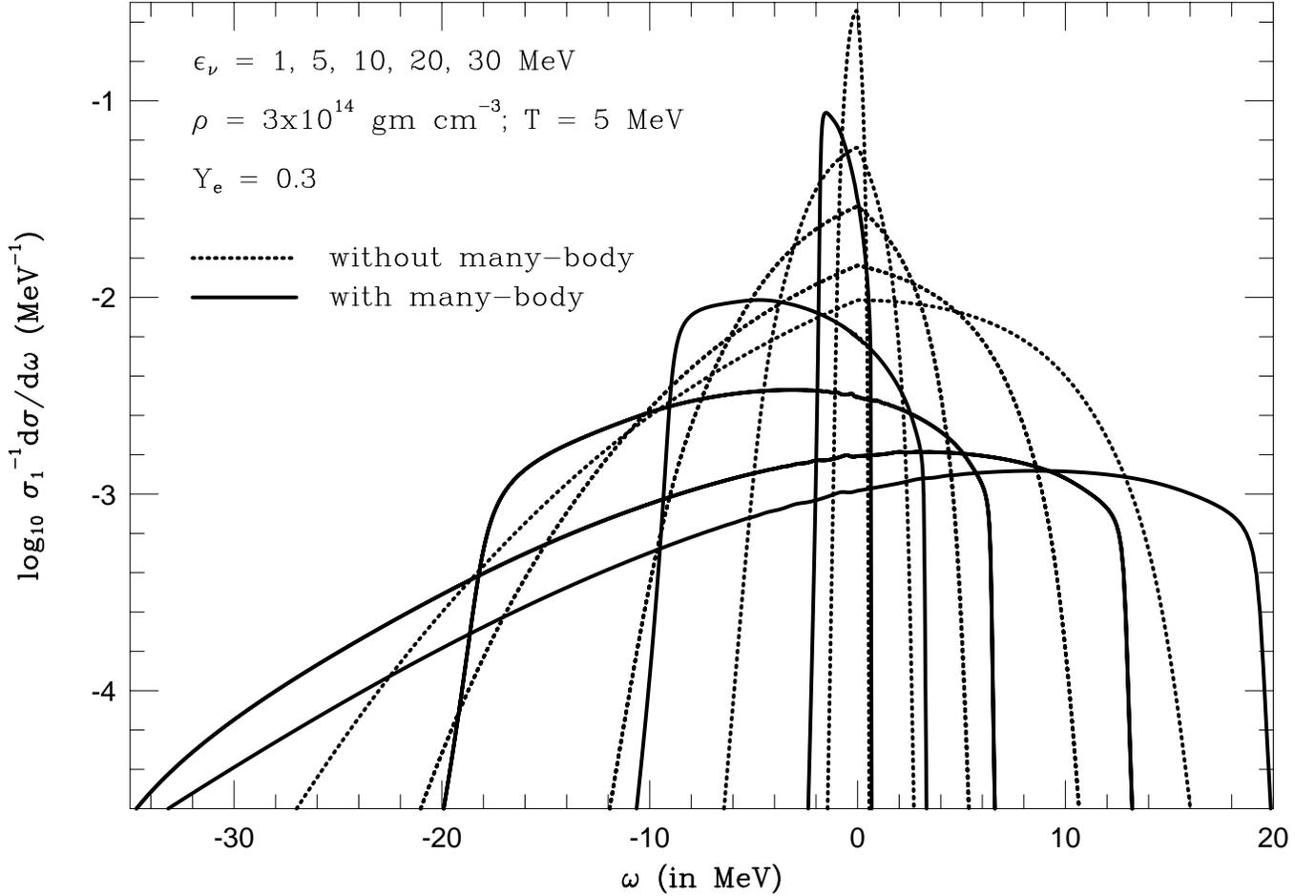}
\end{center}
\vspace*{1.0in}
\caption{The log$_{10}$ of the differential cross section for $\nu$--nucleon scattering versus
the energy transfer, $\omega$, for various values of 
the incident neutrino energy ($\epsilon_\nu $ = 1, 5, 10, 20, 30 MeV).
The dashed curves neglect the many--body effects associated
with $m^*$ and $ \cal C_{V,A} $, while the solid curves include them.  
A density of $3\times10^{14}$ gm cm$^{-3} $, a temperature of 5 MeV,
and an electron fraction, $Y_e$, of 0.3 were assumed. The curves were normalized to the 
total $\nu$--nucleon scattering cross section without nucleon blocking or many--body effects.
}
\label{fig3}
\end{figure}

We depict in Figure~\ref{fig3} the singly--differential 
scattering cross section, divided (normalized) by the total
scattering cross section off of nucleons ($\sigma_1$), for a range
of incident neutrino energies.   We have used eqs.~(\ref{a38}), (\ref{vectel}), 
(\ref{vectel2}), (\ref{axel}), and (\ref{axel2}) to generate
these curves. Figure~\ref{fig3} is a study of the dependence on incident neutrino 
energy (1, 5, 10, 20, 30 MeV), at a fiducial (but 
arbitrary) density ($\rho = 3\times10^{14}$ gm cm$^{-3}$),
temperature ($T$= 5 MeV), and electron fraction ($Y_e$ = 0.3).  
Also shown as dashed lines are the corresponding
curves without the many--body correlation effects 
($\cal C_{V,A}$) and for $m^*=m_n$, but with
final--state nucleon blocking.  Table I depicts the corresponding many--body
suppression factors, obtained by integrating under the curves, as well
as both the average energy transfers ($<\!\omega\!>$) 
and the $rms$ of the energy transfers ($\omega_{rms}$).
Without neutrino blocking, the total suppressions we present in the tables
are very close to the full results for $\nu_\mu$s and $\nu_\tau$s.
For $\nu_e$s, it is more important to include the $ (1-f_{\nu}(E_2)) $ term,
since there is a net electron lepton number in protoneutron stars for most
of their interesting lives. Figure~\ref{fig3} is rich 
with information that we will try to summarize.
Positive $\omega$ represents energy lost from the neutrino and negative $\omega$
represents energy gained by the medium.  First, notice that the widths of the curves
increase with the incident neutrino energy.  Even for the curves without many--body
effects, the widths as depicted in Figure~\ref{fig3} and in Table I are quite large.   
Such widths call into question the elastic approximation \cite{bru2,lp}, but also
imply that energy equilibration for the $\nu_\mu$s and $\nu_\tau$s 
by $\nu$--nucleon scattering 
dominates over the annihilation process, $\nu + \bar{\nu}\rightarrow e^+ + e^-$, and 
$\nu$--electron scattering, both with much lower cross sections \cite{hr}. 
Equilibration for the $\nu_e$s is still via the charged--current absorption
process, $\nu_e + n\rightarrow e^- + p$.

One major reason the widths are larger than are familiar is that in the past
people thought that the neutrino could lose in $\nu$--nucleon scattering 
an energy equal to only about $-E_1^{2}/m_nc^2$,
{\it i.e.} that the fractional energy lost is of order $p_\nu/m_nc$ ($\sim$1\%).
However, this assumes that the nucleons are stationary.  In fact, they are 
thermal and,  the fractional energy they can transfer in a collision to the
neutrino is of order $p_n/m_nc$.  Since the nucleons have such a large mass, if they 
and the neutrino have the same energy, $p_n/m_nc$ is much larger than $p_\nu/m_nc$,
at incident neutrino energies of 10--30 MeV by as much as an order of magnitude.
The formalism we employ incorporates the kinematics of such a collision,
a realistic Fermi--Dirac energy distribution for the nucleons, and final--state nucleon
blocking.  The upshot is the broad distributions, even without the
$\cal C_{V,A}$s ,  seen in Figure~\ref{fig3} and 
tabulated in Table I.  Including many--body effects further 
flattens and broadens the distribution (see below), while lowering the central values of 
${d\sigma}/{d\omega}$, as well as the total integral over $\omega$. 

As Table I demonstrates, even without many--body effects, nucleon blocking
is a large effect, larger than the elastic $\frac{3}{2}\frac{T}{\mu_n}$ correction
that comes from the low--temperature expansion in powers of $T$.
That correction can be applied in the extreme degenerate limit,
but the nucleons are only partially degenerate in protoneutron stars 
and supernovae, even at nuclear densities, and the extreme degenerate limit is never 
achieved.  The error in using the $\frac{3}{2}\frac{T}{\mu_n}$--correction ansatz 
can be as much as a factor of two, depending upon the nuclear interaction model.

Note in Table I that the many--body cross section suppression factors
are not strong functions of the incident neutrino energy, except at very low
energies ({\it e.g.}, 1 -- 5 MeV), but that the suppression factors themselves
are quite large at this fiducial thermodynamic point (near nuclear density).  
The extra suppression, beyond that due to nucleon blocking, 
is due both to the decrease in the effective mass (factor of $\sim$1.5)
and to the many--body correlation terms, $\cal C_{V,A}$ .
The total factor is between 10 and 30.  This means that the combination 
of final--state nucleon blocking (a factor of $\sim$6 by itself) and many--body
effects (another factor of 3 to 4) renders the supernova core much
more transparent to neutrinos than previously thought.  We remind the reader
that we are here calculating only the neutral--current rates.  While they
dominate for the $\nu_\mu$s and $\nu_\tau$s, the charged--current dominates
for the $\nu_e$s.  Nevertheless, we are in the midst of preliminary calculations 
that indicate that the suppression of these rates is also quite large
(Burrows \& Sawyer, in preparation; see also Reddy {\it et al.}, in preparation).

The previous assumptions concerning the distribution
in energy transfers for $\nu$--nucleon scattering were more akin to the narrow
(quasi--delta function) curve, seen in Figure~\ref{fig3} 
for 1--MeV incident neutrino energy. 
The contrast between that curve and the others is manifest.

\begin{figure}
\begin{center}
\leavevmode
\epsfysize=1.00\hsize
\epsfbox[160     160   612   792]{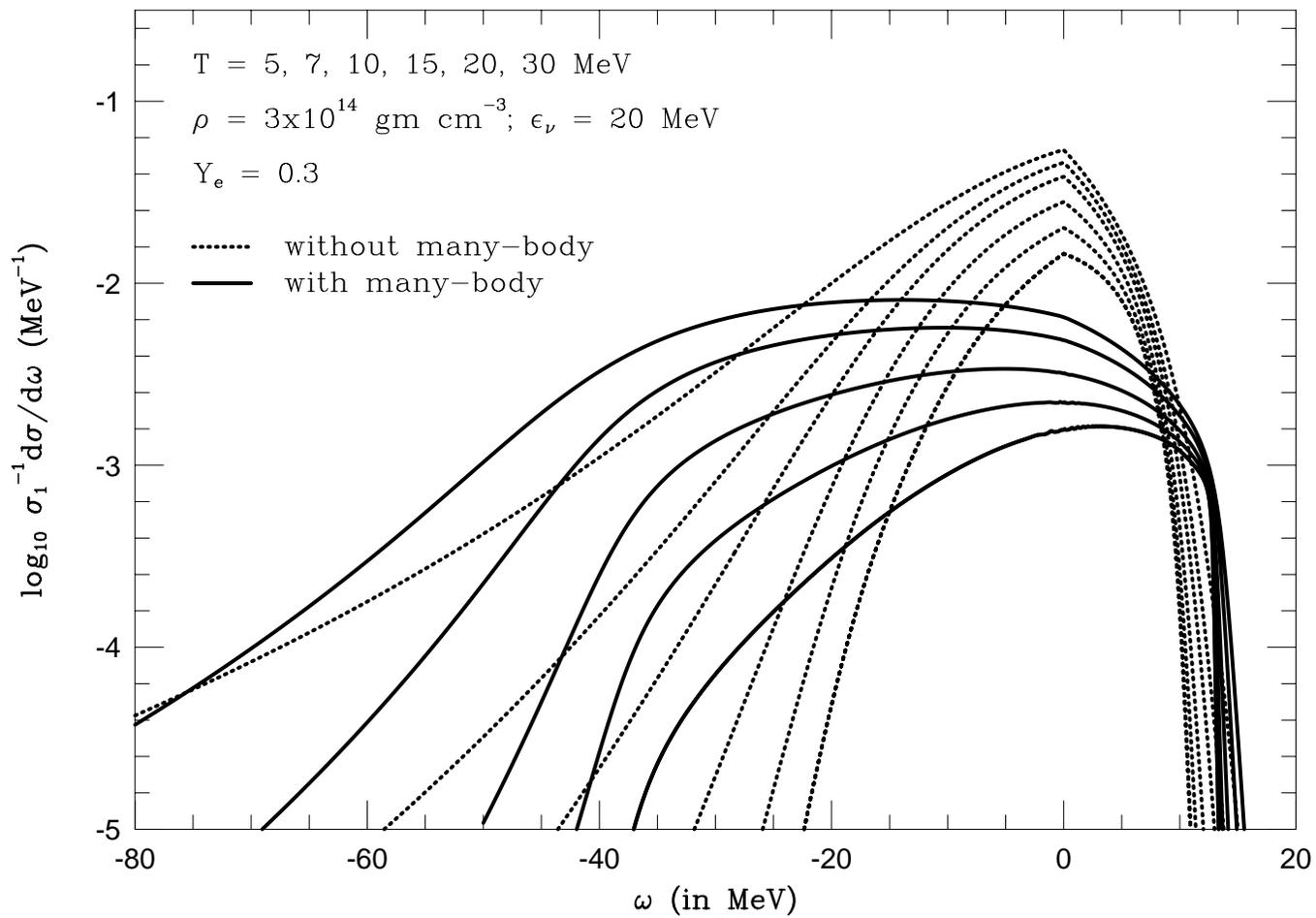}
\end{center}
\vspace*{2.0in}
\caption{Similar to Figure 3, but for various temperatures (5, 7, 10, 15, 20, 30 MeV)
and at an incident neutrino energy of 20 MeV.
}
\label{fig4}
\end{figure}

Figure~\ref{fig4} is similar to Figure~\ref{fig3}, but for a range of
temperatures (5, 7, 10, 15, 20, 30 MeV).  The calculations were done at $E_1 = 20\, {\rm MeV}$.
Superposed are the corresponding reference curves for $\cal C_{V,A}$ = 1 and $m^*=m_n$.
Table II lists the suppression factors, $<\!\omega\!>$, and $\omega_{rms}$.
As the temperature rises, the suppression diminishes.  
It is clear in Figure~\ref{fig4} that energy transfer from the medium to 
the neutrino becomes more likely as the temperature rises.  This is to be 
expected and is all the more pronounced in the many--body case.
As in Figure~\ref{fig3}, the differential cross section is flattened when interaction
effects are included, but Figure~\ref{fig4} shows that this flattening effect
is more significant at higher temperatures.  As Table 
II demonstrates, though the total suppression
diminishes with temperature, many--body effects still increase it, the more so at lower 
temperatures.  Given the large values of $<\!\omega\!>$ and $\omega_{rms}$, 
it is difficult to see how even a Fokker--Planck treatment 
of neutrino energy redistribution via these processes could be viable
and the full redistribution formalism \cite{mezz,tubbs} may be required, at least for the 
$\nu_\mu$s and the $\nu_\tau$s.

\begin{figure}
\begin{center}
\leavevmode
\epsfysize=1.00\hsize
\epsfbox[160     160   612   792]{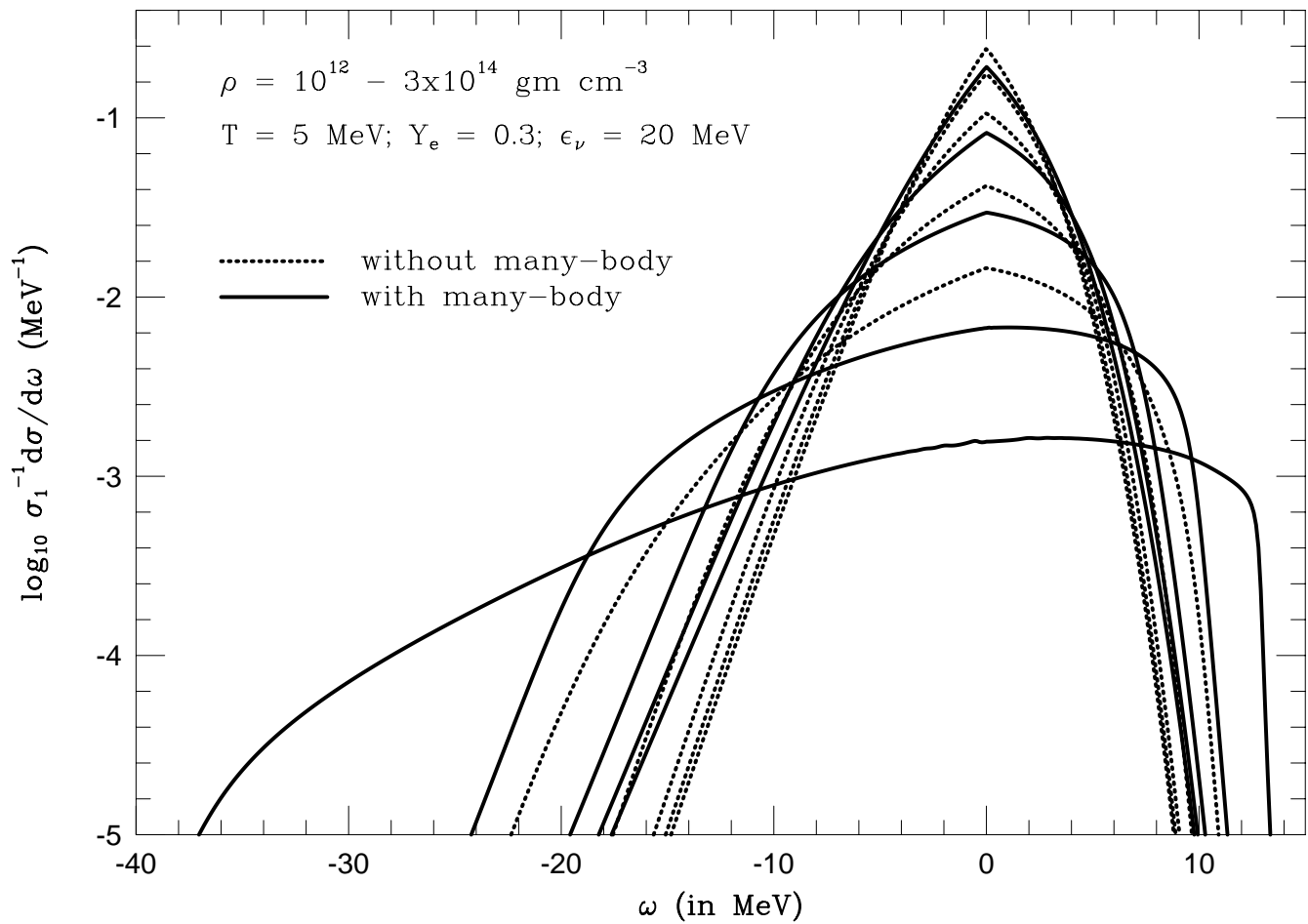}
\end{center}
\vspace*{2.0in}
\caption{Similar to Figure 3, but for various densities ($10^{12}$, $10^{13}$,
$3\times10^{13}$, $10^{14}$, and $3\times10^{14}$  gm cm$^{-3}$), a temperature
of 5 MeV, and an incident neutrino energy of 20 MeV.
}
\label{fig5}
\end{figure}

Figure~\ref{fig5} depicts the density dependence of the singly--differential cross section
from $\rho = 10^{12}$ gm cm$^{-3}$ to $\rho = 3\times10^{14}$ gm cm$^{-3}$, with and without
many--body effects.  The temperature is kept constant at 5 MeV and the incident 
neutrino energy is 20 MeV.  Table III displays the corresponding total suppression
factors, $<\!\omega\!>$, and $\omega_{rms}$.   Both Figure~\ref{fig5} and Table III
demonstrate that the total suppression effect increases quickly with density 
beyond 10$^{13}$ gm cm$^{-3}$ and that that due solely to many--body effects increases quickly
beyond 10$^{14}$ gm cm$^{-3}$.  In fact, the total correlation suppression factor reaches 
$\sim$100 at 10$^{15}$ gm cm$^{-3}$, calling into question the perturbative assumption itself.
Nevertheless, it is clear that correlation and many--body effects can drastically lower
the neutral--current rates in high--density supernova cores.  Note that the widths
displayed in Figure~\ref{fig5} increase with increasing $\rho$. This reflects the increasing
nucleon degeneracy and increasing average nucleon energy that accompanys increasing $\rho$.

Taking a temperature, density and $Y_e$ profile from an early post--bounce model
of Burrows, Hayes, \& Fryxell \cite{bhf}, we show in Figure~\ref{fig6}  
the corresponding differential cross section curves
at $E_1= 20$ MeV for a ``realistic" profile.  Table IV displays the total suppression
factors, $<\!\omega\!>$, and $\omega_{rms}$.  The total suppression effect at the center
is greater than 30, that due solely to many--body effects is $\sim$5. 
The $<\!\omega\!>$ and $\omega_{rms}$ are also large.  Even at $10^{14}$ gm cm$^{-3}$, the 
extra suppression effect due to many--body effects is $\sim$2.

It is interesting to compare the suppression factors calculated from
our full equations with those that come from the long--wavelength
limits given in eqs.~(\ref{sv43}) and (\ref{sa45}), 
using the same potential parameters.
At a density of $3.0 \times 10^{14}$ gm cm$^{-3}$,  a
temperature of 5 MeV, and for neutrino energies less than 10 MeV,
the factors, calculated the two different
ways, are within 10\% of each other.
However, for a neutrino energy of 30 MeV, the full calculation
gives a suppression that is 30\% greater than
in the long--wavelength limit. It will be 
recalled that our ``long--wavelength'' limit
also requires that the energy transfer, $\omega$, be small.
Since the energy transfers are of the order of
$p_F q$, this limit can be achieved with fixed nucleon mass only by going
to lower density.  With temperature fixed at 5 MeV, and neutrino energy
fixed at 20 MeV, we find less than a 5\% discrepancy for a density
of $10^{13}$ gm cm$^{-3}$. However, the discrepancy 
rises steadily to nearly 50\% at a density of $10^{14}$
gm cm$^{-3}$. These numbers confirm that the static
approach is adequate only in some of parameter realms occupied
by supernovae.
  
\subsection{Correction Factors versus Energy Transfer}

\begin{figure}
\begin{center}
\leavevmode
\epsfysize=1.00\hsize
\epsfbox[160     160   612   792]{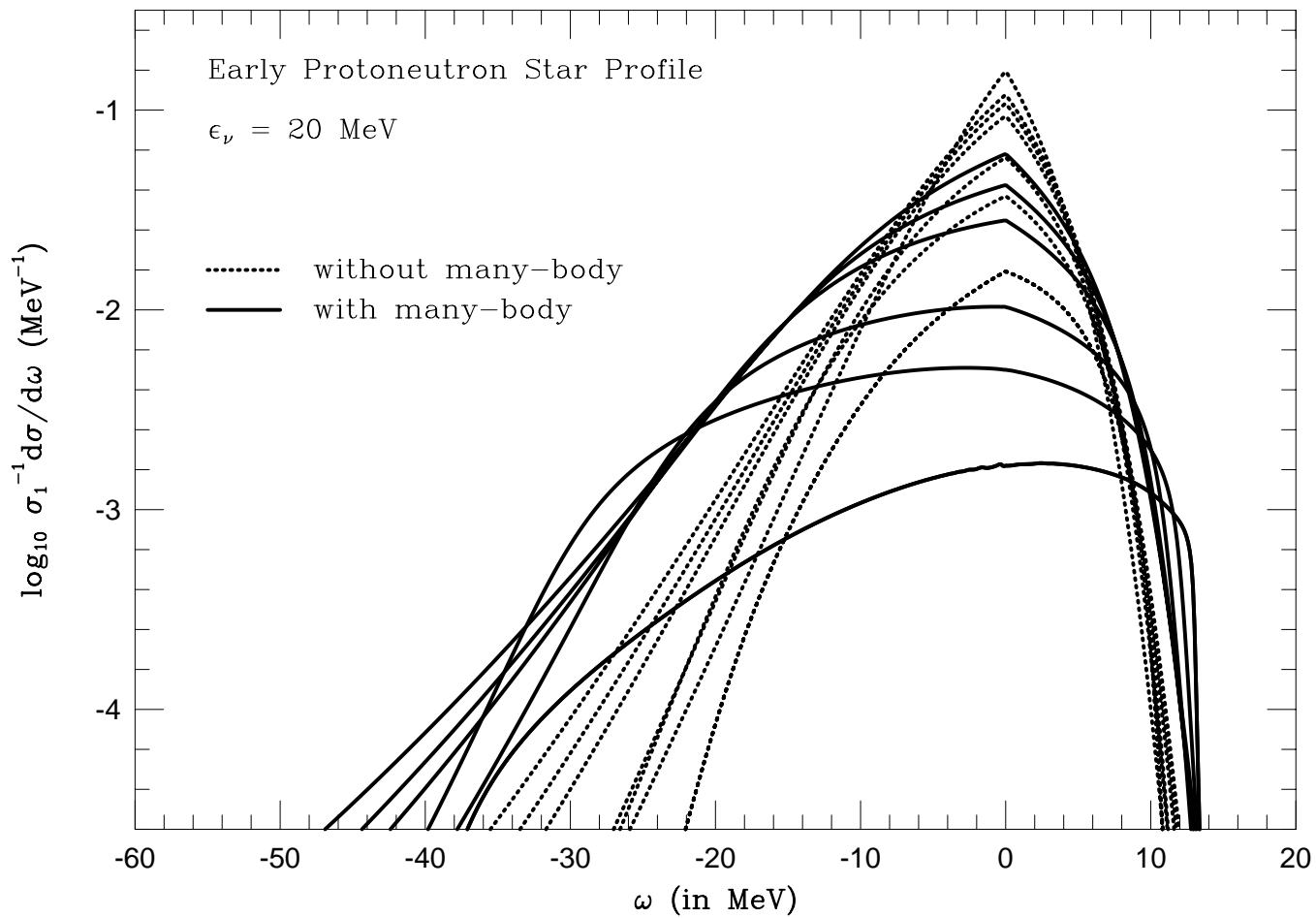}
\end{center}
\vspace*{2.0in}
\caption{Similar to Figure 3, but for an early post--bounce temperature-density-$Y_e$ profile
from Burrows, Hayes, \& Fryxell (1995) (see Table IV).  
The incident neutrino energy was assumed to be 20 MeV.
}
\label{fig6}
\end{figure}

\begin{figure}
\begin{center}
\leavevmode
\epsfysize=1.00\hsize
\epsfbox[160     160   612   792]{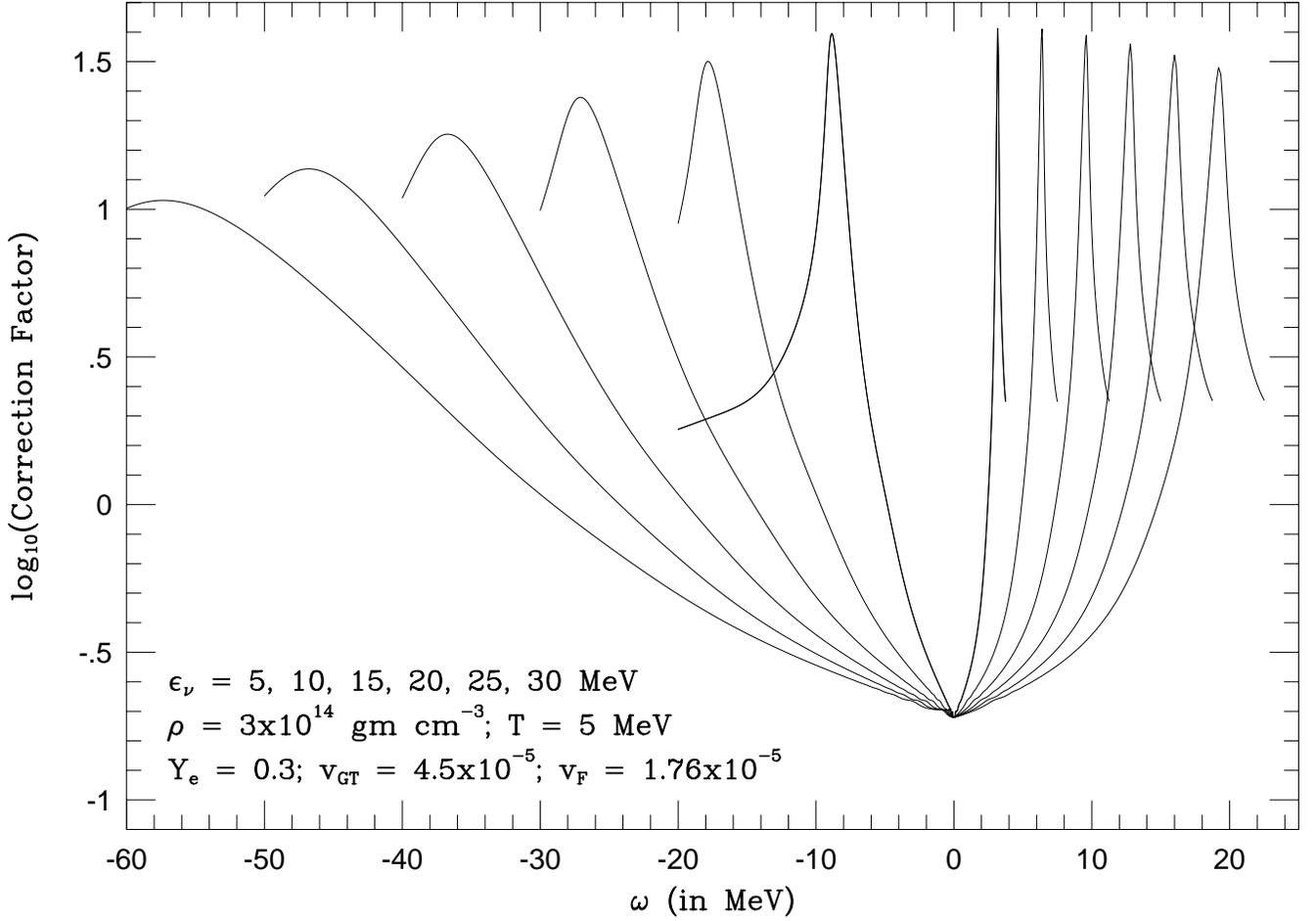}
\end{center}
\vspace*{1.5in}
\caption{Log$_{10}$ of the correction factor due to both $\cal C_{V,A}$ and an effective
mass of 0.75$m_n$ that corresponds to Figure 3 (the $\epsilon_{\nu} = 1$ MeV
line has been omitted).  There is suppression at low $|\omega|$s, but
a resonant enhancement at high $|\omega|$s.  Note that at very small $|\omega|$s the factor
is a weak function of incident neutrino energy.
}
\label{fig7}
\end{figure}

\begin{figure}
\begin{center}
\leavevmode
\epsfysize=1.00\hsize
\epsfbox[160     160   612   792]{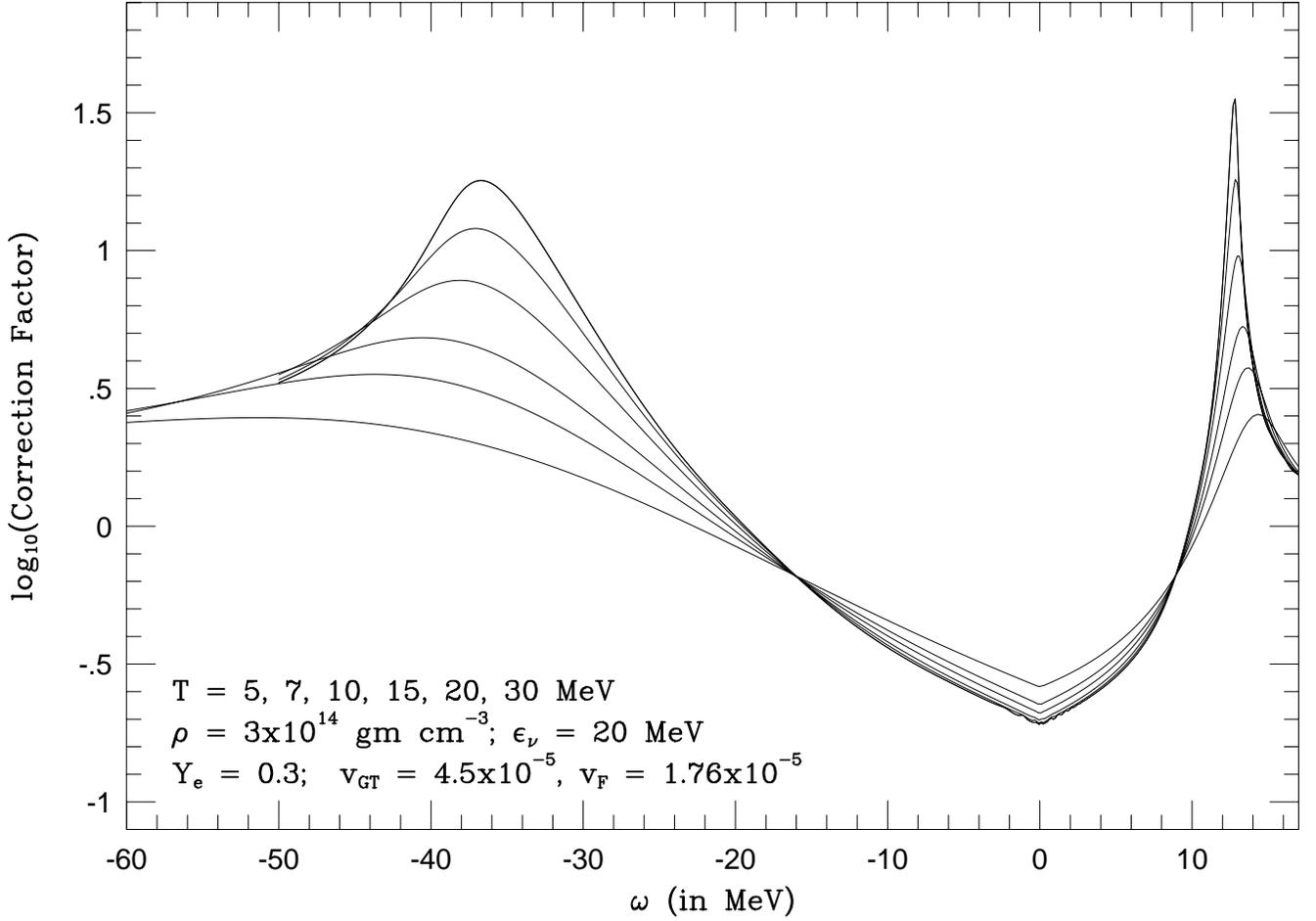}
\end{center}
\vspace*{1.5in}
\caption{Log$_{10}$ of the correction factor due to both $\cal C_{V,A}$ and an effective
mass of 0.75$m_n$ that corresponds to Figure 4.  There is suppression at low $|\omega|$s, but
a resonant enhancement at high $|\omega|$s. Note that at small $|\omega|$s the 
magnitude of the correction is a decreasing function of temperature.
}
\label{fig8}
\end{figure}

\begin{figure}
\begin{center}
\leavevmode
\epsfysize=1.00\hsize
\epsfbox[160     160   612   792]{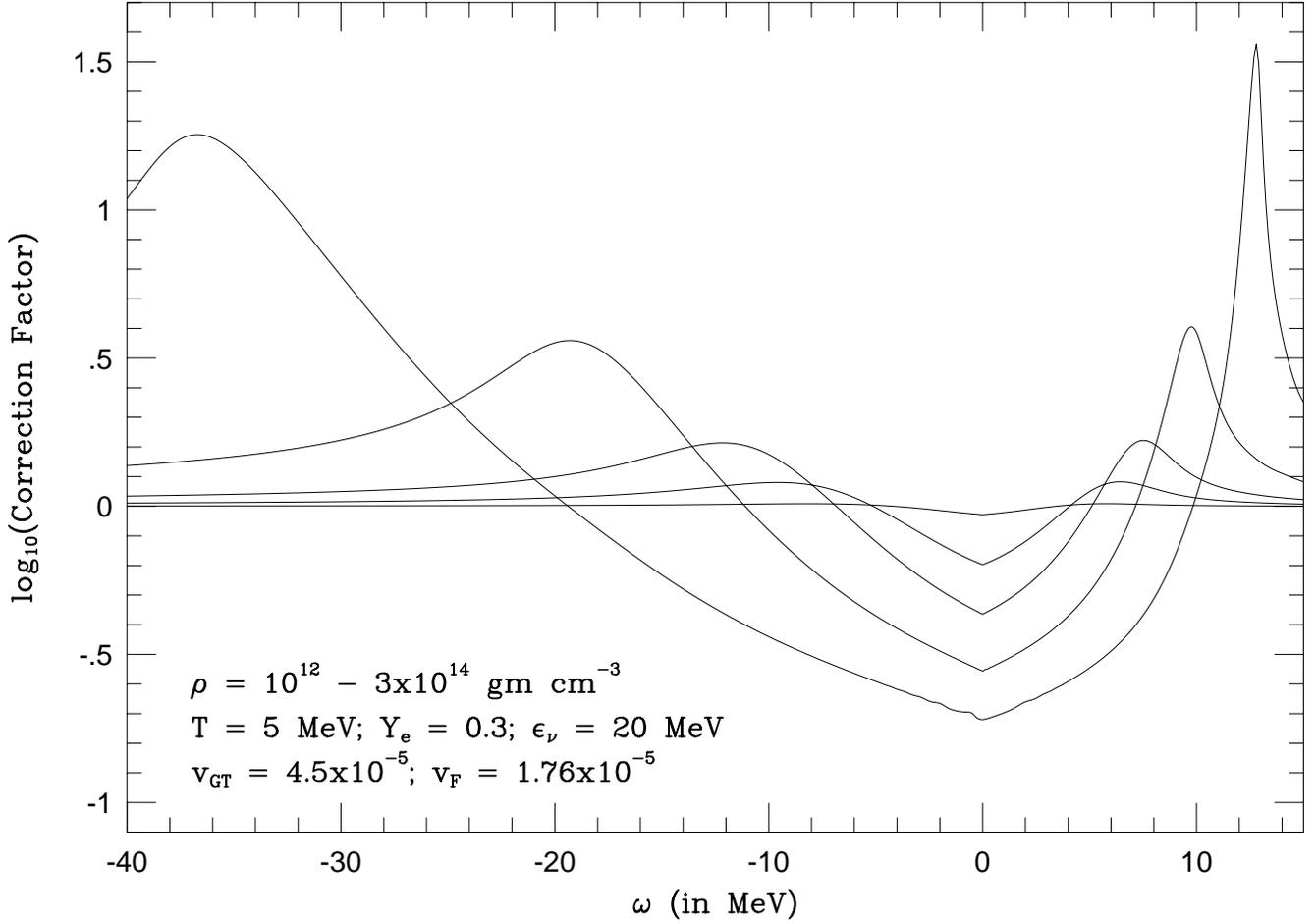}
\end{center}
\vspace*{1.5in}
\caption{Log$_{10}$ of the correction factor due to both $\cal C_{V,A}$ and an effective
mass of 0.75$m_n$ that corresponds to Figure 5.  There is suppression at low $|\omega|$s, but
a resonant enhancement at high $|\omega|$s. Note that at small $|\omega|$s the factor
is a strong function of density.
}
\label{fig9}
\end{figure}

\begin{figure}
\begin{center}
\leavevmode
\epsfysize=1.00\hsize
\epsfbox[160     160   612   792]{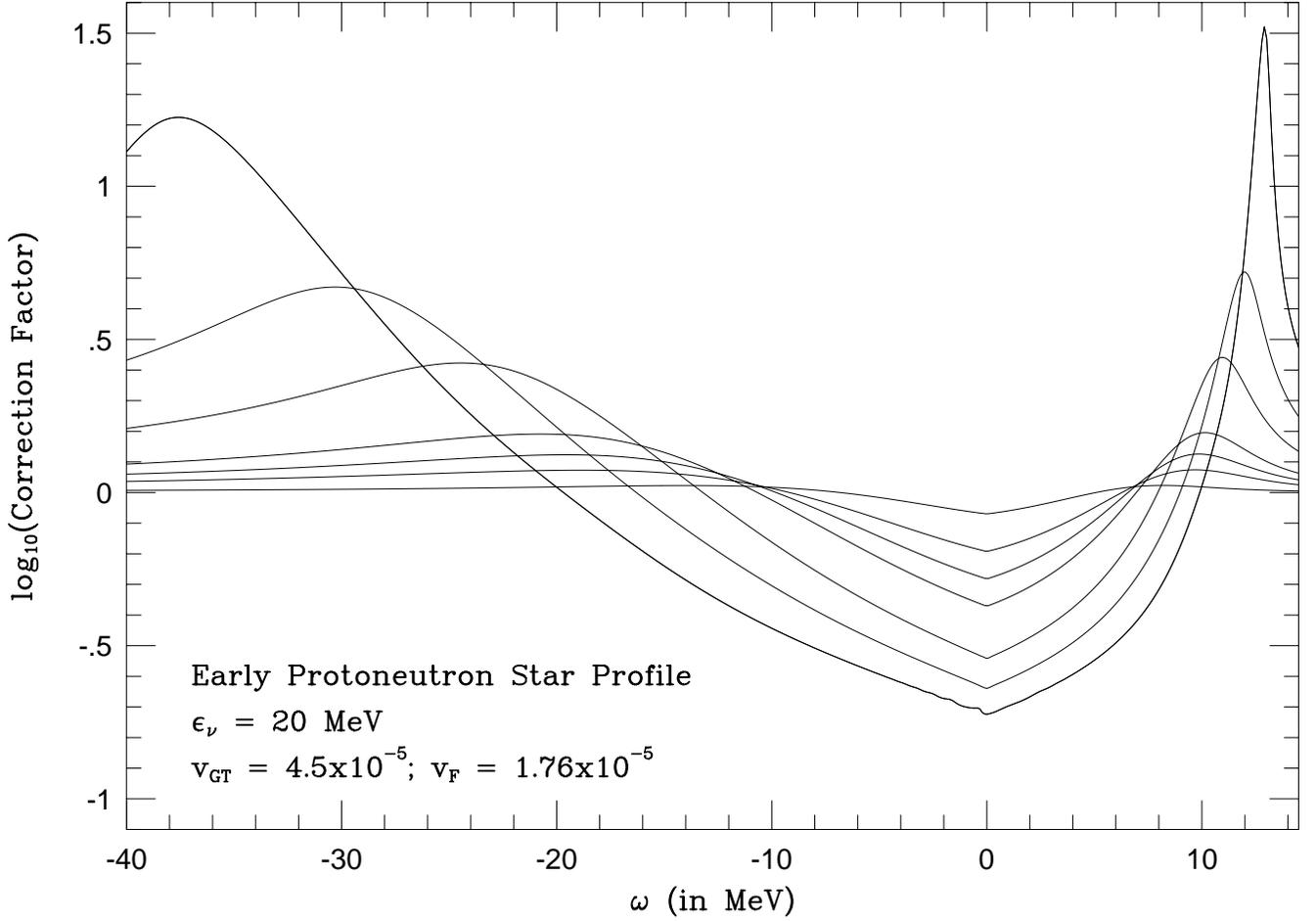}
\end{center}
\vspace*{1.5in}
\caption{Log$_{10}$ of the correction factor due to both $\cal C_{V,A}$ and an effective
mass of 0.75$m_n$ that corresponds to Figure 6.  There is suppression at low $|\omega|$s, but
a resonant enhancement at high $|\omega|$s. Note that at small $|\omega|$s the magnitude of
the correction is a strong function of position in the star and is largest at the center.
}
\label{fig10}
\end{figure}

It is instructive to calculate $d^2 \Gamma/d \omega \hspace{1 pt} d \cos\theta$, 
after integration over $\cos\theta$,  ({\it i.e.}, ${d\sigma}/{d\omega}$), with and without  
the $\cal C_{V}$ and $\cal C_{A}$ terms and a renormalized nucleon mass,
and to take the ratio. This gives one a sense of the integrated 
correction factor to the differental rate due to many--body effects.
This is the many--body correction factor to what might be called $S(\omega)$.
Figures~\ref{fig7}, \ref{fig8}, \ref{fig9}, and \ref{fig10} depict the logarithms of
these correction factors as a function of $\omega$.  These figures correspond to the
Figures~\ref{fig3} through \ref{fig6}.  We see the expected suppression factor
in the small--$\omega$ regime that dominates the total suppression integral,
but at high $|\omega|$s we see  manifestations of the resonances.  The rapid decrease in 
${\rm Im}\Pi^{(0)}(q,\omega)$ with $|\omega|$ defeats the increase in the 
correction factor at high $|\omega|$ in the total cross section integral.
However, this resonance effect is partly responsible for the high--$|\omega|$
flattening we see in Figures~\ref{fig3} through \ref{fig6}.
Note that this correction factor does not go to zero at $\omega=0$, can go
far above 1 for high $|\omega|$s, and does not have the form, $\omega^2/(\omega^2 + K)$.  

\section{An Illustrative Model of Protoneutron Star Cooling}

While the consequences of the inclusion of many--body 
effects on neutrino cross sections and energy redistribution
are by no means clear, one can straightforewardly ascertain the potential
of such effects to alter aspects of supernova and protoneutron star development
by performing simplified protoneutron star cooling calculations.
We have chosen to investigative the evolution of the electron neutrino luminosity
($L_{\nu_e}$), under certain simplifying assumptions, with
and without various ad hoc alterations to the neutrino opacities.  The code,
developed by Burrows \& Lattimer \cite{bl}, uses the Henyey technique with a simple
nuclear EOS, is general--relativistic, and handles the transport of neutrinos
of all species in the diffusion approximation.   It conserves total energy to about 
one percent over 20 seconds of evolution. The default cases are models of
$L_{\nu_e}$ versus time after bounce without accretion and with an assumed accretion
rate of $0.4 {\rm M}_{\odot}\, e^{-t/\tau_a}$, where $\tau_a$ is 0.5 seconds. The initial
baryon mass is 1.3\, ${\rm M}_{\odot}$.  We alter the total opacities in an ad hoc fashion  
to mimic the decreases we can anticipate given our preliminary exploration
of neutral--current suppressions and compare the results to the default models.   
The two exploratory models we present assume that the total opacities
are decreased by a fixed amount above a given density, in one case $5\times10^{13}$ 
gm cm$^{-3}$ and in the other $10^{14}$ gm cm$^{-3}$.  For the $5\times10^{13}$
gm cm$^{-3}$ case the suppressions were 0.3, 0.1, and 0.05 (extreme) and for the $10^{14}$ gm cm$^{-3}$
case the suppressions were 0.3 and 0.1.  These were guided by our results,
but should be considered arbitrary.  Note that the more conservative case
for which the opacity is altered modestly and at the higher density is the more
likely.

\begin{figure}
\vspace*{-1.15in}
\begin{center}
\leavevmode
\epsfysize=1.00\hsize
\epsfbox[160     160   612   792]{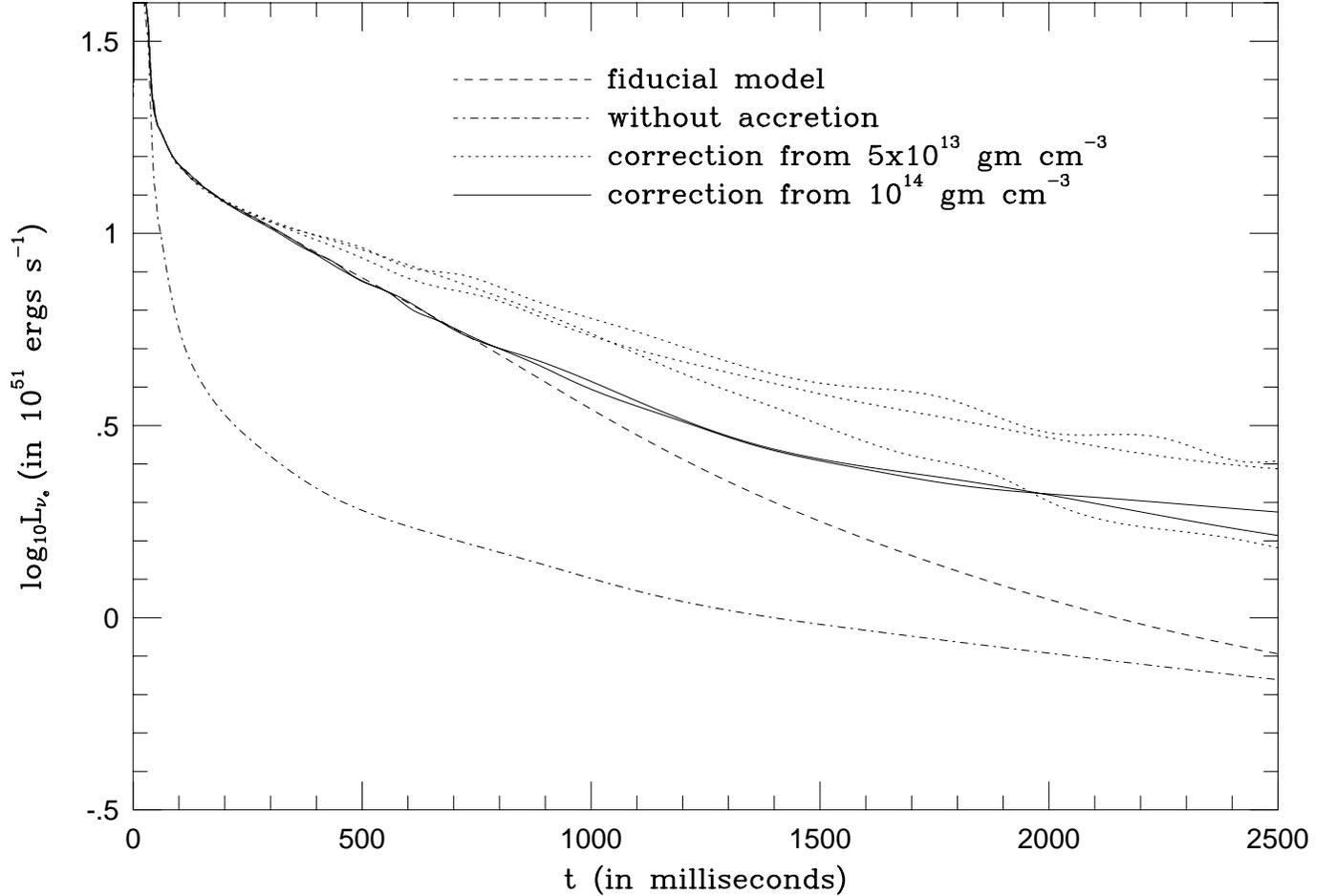}
\end{center}
\vspace*{0.93in}
\caption{Log$_{10}$ of the electron neutrino luminosity ($L_{\nu_e}$) in units of 
$10^{51}$ ergs s$^{-1}$ versus time after bounce in milliseconds, with and without accretion.
For the accretion models, total opacity suppression factors of 0.3, 0.1, and 0.05 were
assumed above $5\times10^{13}$ gm cm$^{-3}$ and of 0.3 and 0.1 were assumed above $10^{14}$ gm cm$^{-3}$.  
The fiducial model is dashed, the model without accretion is dot--dashed, the models
with correction above $5\times10^{13}$ gm cm$^{-3}$ are dotted, and those with correction above
$10^{14}$ gm cm$^{-3}$ are solid.  On this plot, the models with the largest corrections
have the highest luminosities after 2500 milliseconds.  The comparisons between the dashed
curve and all others are the most germane.
}
\label{fig11}
\end{figure}

The results are plotted in Figure~\ref{fig11}.   From a comparison
between the fiducial model and the one without accretion,  one notes that accretion 
dominates as a power source in the early seconds.   If the supernova is reignited
within the first hundreds of milliseconds, it is unlikely that the new opacities
at high densities will play a central role in its revitalization.  However, if the delay
to explosion is many hundreds of milliseconds to seconds, higher $L_{\nu_e}$s occasioned 
by the many--body and final--state nucleon blocking effects may well play a role
in the supernova, and in powering the explosion after it is relaunched.
Even partially enhanced luminosities have been shown to be important for neutrino--driven 
explosions \cite{bhf,mw,janka1,bg}.   As Figure~\ref{fig11} demonstrates,
the lower the density at which opacities are altered, the earlier the effect.
The enhancements in the luminosities after a few seconds can be a factor of 
two, though enhancements of 30\%--50\% are more likely.  Since energy
is conserved, the luminosities at later times (5--60 seconds, not shown) are decreased
relative to the those of the fiducial model.  Within this model set, the theoretical signals in 
Kamioka II \cite{hir} and IMB \cite{bion} range by no more than a factor of two. 
Drastic decreases in the total opacities translate into only
moderate (but intriguing) changes in $L_{\nu_e}$ and the signals because
even a completely transparent core merely flattens the interior temperature gradients.
The energy must still diffuse through the unaltered density region
below $5\times10^{13}$ or $10^{14}$ gm cm$^{-3}$.  

We point out that for progenitor stars in the $8\, {\rm M}_{\odot}$  to $13\, {\rm M}_{\odot}$
mass range early accretion onto the protoneutron star before explosion
will be slight.   Hence, alterations in the high--density neutrino cross sections 
might show themselves earlier in lower--mass progenitors, 
in time to be more unambiguously of importance 
in reigniting and powering the supernova.  Nevertheless, the reader should be cautioned
that these conclusions await a more thoughtful implementation of these new neutrino opacities
and that the calculations depicted in Figure~\ref{fig11} are merely suggestive.

\section{Conclusions}

In this paper, we have developed a consistent formalism for calculating 
the Fermi and Gamow--Teller structure functions in nuclear matter,
including many--body correlations and the full kinematics 
of neutrino--nucleon scattering.  Previously all but ignored, the 
effects of density--density and spin--spin correlations on
neutrino scattering rates are found to be significant.
Above nuclear density, the total cross section suppression factor 
due to final--state nucleon blocking, correlations, and
a reduction in the nucleon effective mass can be more than an 
order of magnitude.  The upshot is that supernova cores are
more transparent than previously thought.  Since the many--body
corrections to the charged--current rates for the $\nu_e$s 
have yet to be published, in this paper we conclude only
that the total opacities of nuclear matter to $\nu_\mu$s and
$\nu_\tau$s (and their anti--particles) are qualitatively altered.
The magnitude of the rate corrections due to many--body effects
increases with density, decreases with temperature,
and is roughly independent of incident neutrino energy.

In addition, we find that the neutrino--matter energy transfer rates
due to neutrino--nucleon scattering are much larger than heretofore
assumed.  As a consequence,  energy equilibration for the
non--electron--type neutrino species may well be by what was
previously considered an elastic process.  This obtains even for
free nucleons at high densities, but is all the more true with
many--body effects included. An identifiable part of the energy
transfer enhancement comes from the excitation of collective
modes in the matter, modes akin to the Gamow--Teller and Giant
Dipole resonances in nuclei. Energy is transferred by the
\v{C}erenkov mechanism to or from these modes when
the energy transfer ($\pm\omega$) and momentum transfer ($q$)
satisfy the dispersion relation of the
medium's excitations (zero sound and spin sound, both of which
require repulsive interactions).

Under the assumption that both the charged--current and the neutral--
current cross sections are decreased, we calculated a set of ad hoc protoneutron 
star cooling models to gauge the importance of the new opacities 
to the supernova itself.  While the early luminosities are 
not altered, the luminosities after many hundreds of milliseconds to 
seconds can be altered by factors that range from 10\% to 100\%.     
Such factors  may have a bearing on the efficacy of the
neutrino--driven supernova mechanism, the delay 
to explosion, the energy of the explosion,
and the strength and relative role of convective overturn.  
The magnitude of the relative enhancement of the driving luminosity
is a function of the post--bounce mass accretion rates and may be larger 
for the least massive, massive stars.  Hence, these new opacities
may be more germane to the terminal behavior of the massive stars 
most favored by the IMF.

It remains to derive the effects of correlations on the charged--current
rates, as well as to develop a simple algorithm for incorporating
these effects into supernova and protoneutron star codes, 
with full energy redistribution.  Since many--body effects at the  
highest densities are quite large, a non--perturbative approach
may need to be developed for the deep interiors.
Furthermore, the structure function calculations 
should be done in a manner fully consistent with the
equation of state employed, since the physics
of the two are inextricably linked.  

If we find that there is indeed an across--the--board reduction in the 
neutrino--matter rates at high densities, it will be yet another reminder
that the keys to the supernova puzzle and its systematics 
lie not in any one realm of expertise (radiative transfer, hydrodynamics,
the equation of state, the weak interaction, etc.), but in all,  and that the 20\%
effects we seek under lamp posts might at times divert us from discovering 
those far larger just a few steps down the street.

\section*{ACKNOWLEDGMENTS}

We would like to acknowledge stimulating and productive
conversations with S. Reddy, M. Prakash, 
G. Raffelt, D. Seckel, and J. Lattimer,
as well as support from the NSF under grant No.
AST-96-17494.  We would also like to acknowledge
the germinating influence of the Santa Barbara
Institute for Theoretical Physics, supported
by the NSF under grant No. PHY94-07194.

\section*{Appendix A}

To get expressions for the real and imaginary parts of $\Pi^{(0)}$ 
that are efficient for computation, we begin with FW (33.4), with 
notations changed to those of this paper

\begin{equation}
\Pi^{(0)}(q,\omega)=-2 \int\frac{d^3p}{(2\pi)^3}\frac{f_{\bf p+q}
\it-f_{\bf p\it}}{\omega+i\eta+\epsilon_{\bf p}-\epsilon_{\bf p+q}}.
\end{equation}

We can write each of the Fermi--Dirac distributions that 
appear here as a sum of Boltzmann distributions,

\begin{equation}
f(\bf p \it)=\sum^{\infty}_{j=1}(-1)^{j+1}e^{j\beta(\mu-\epsilon_p)}.
\end{equation}

We use this expansion, together with FW (33.8) for the imaginary 
part of $\Pi^{(0)}$ in the Boltzmann limit, to obtain,

\begin{equation}
{\rm Im} \Pi^{(0)}(q,\omega)=\frac{m^2}{2 \pi \beta q}\sum^{\infty}_{j=1}
(-1)^{j+1}j^{-1}\Bigl[e^{j(\beta\mu-Q^2)}
-e^{j(\beta\mu-Q^2-\beta\omega)}\Bigr],
\label{46}
\end{equation}
where 

\begin{equation}
Q=\bigl(\frac{m\beta}{2}\bigr)^{1/2} \bigl(-\frac{\omega}{q}+\frac{q}{2m} \bigr)
\end{equation}
and the terms with $j>1$ in the series come from the replacement, 
$\beta\rightarrow\ j\beta$, in the Boltzmann result.
The sum is easily performed to give the result shown in (\ref{a25}).
For the real part, we use the same trick, beginning with the Boltzmann form (FW 33.9),

\begin{equation}
{\rm Re} \Pi^{(0)}(q,\omega)_{boltz}=\frac{m^2}{2\pi^{3/2}\beta q} 
e^{-\beta \mu}\Phi(Q)+(\omega\rightarrow -\omega),
\end {equation}
where 

\begin{equation}
\Phi(Q)=2Q\int_0^1dy\, e^{Q^2(y^2-1)}.
\end{equation}

We obtain,

\begin{equation}
{\rm Re} \Pi^{(0)}(q,\omega)=\frac{m^2}{2\pi^{3/2}\beta q}Q\int_0^1dy 
\sum^{\infty}_{j=1}(-1)^{j+1}j^{-1/2}e^{j(\beta\mu+Q^2(y^2-1))} + (\omega\rightarrow -\omega).
\end{equation}

Next, we represent the $j^{-1/2}$ factor under the sum by $\pi^{-1/2}
\int_{-\infty}^{\infty}ds \;e^{-js^2}$, so that we can sum the 
geometric series. Displacing the $s$ integration variable $s\rightarrow s+QY$, we obtain, 

\begin{equation}
{\rm Re} \Pi^{(0)}(q,\omega)=\frac{m^2}{\pi^{2}\beta q} Q \int_{-\infty}^{\infty} ds\int_0^1dy 
\frac{e^{-(s^2+2sQy+Q^2+\beta\mu)}}{1+e^{-(s^2+2sQy+Q^2+\beta\mu)}}+(\omega\rightarrow -\omega).
\end{equation}

The $y$ integral is now easily performed to give the result (\ref{a24}).

\section*{Appendix B}

We show directly that in the large--$m$, small--$q$ limits the complete 
correlation functions as calculated in the ring sum in \S IV
reduce to the long--wavelength limits that were determined from 
the single--particle energies in \S II. In the large--$m$ limit, 
the function ${\rm Im}\Pi^{0}(q,\omega)$ becomes more and more concentrated 
around $\omega=0$, so that to calculate the rate we set $\omega$ 
equal to zero, except in the factor,

\begin{equation}
I_1=\pi^{-1}\int_{-\infty}^{\infty} d\omega\, {\rm Im}\Pi(q,\omega)(1-e^{-\beta\omega})^{-1}.
\end{equation} 

In the limit, we make the replacement $(1-e^{-\beta\omega})\rightarrow\beta\omega$ 
and use the Kramers--Kronig dispersion relation in the variable $\omega$ to write
$I_1 \rightarrow  {\rm Re}\Pi(q,\omega=0)$. Finally, we use (\ref{a24}) 
in the limit $Q\rightarrow 0$ to show that ${\rm Re}\Pi^{(0)}(0,0)=-h(\mu)$, where $h(\mu)$
is the function defined in (\ref{a17}). The ring approximation 
for the one--channel case (\ref{a23}) with a potential $U$ now reads,

\begin{equation} 
I_1\rightarrow h(\mu)/(1+Uh(\mu)),
\end{equation}
in agreement with (\ref{b20}).

\newpage

\begin{table}
 \caption{The total suppression factors ($\cal S$), average energy transfers ($<\!\omega\!>$)
and $rms$ energy transfers ($\omega_{rms}$) for various incident neutrino energies
($E_1$).  The temperature is 5 MeV, the density is $3\times10^{14}$ gm cm$^{-3}$,
and the electron fraction is 0.3.  These quantities are shown for both the
many--body case and for the case without correlations 
or a renormalized mass (subscript or superscript $0$). 
All the energies are in MeV.
For the many--body case, the default nuclear model described in the text is employed.
The suppression factors are the factors by which the default
total cross section should be multiplied to obtain the corrected
total cross section.
}
\begin{center}
 \begin{tabular}{ccccccc}
$E_1$ (MeV)  & $\cal S$& $\cal S_{\it 0}$ & $<\!\omega\!>$ (MeV) & $<\!\omega\!>_0$ (MeV)&
$\omega_{rms}$ (MeV) & $\omega_{rms}^0$ (MeV) \\
\hline
1 & 0.131& 0.217& -0.917& -0.207& 0.592& 0.295\\
5 & 0.090 & 0.202& -3.749& -0.852& 3.088& 1.432\\
10 & 0.060& 0.186& -5.064& -3.000& 6.068& 2.755\\
15 & 0.045& 0.173& -4.204& -1.288& 8.241& 3.939\\
20 & 0.038& 0.164& -2.156& -0.970& 9.492& 4.977\\
25 & 0.035& 0.158&  0.287& -0.385& 10.195& 5.881\\
30 & 0.034& 0.154&  2.779&  0.402& 10.687& 6.671\\
40 & 0.035& 0.152&  7.583&  2.375& 11.569& 8.007\\
50 & 0.038& 0.155&  12.109& 4.649& 12.522& 9.143\\
60 & 0.041& 0.161&  16.413& 7.063& 13.574& 10.177\\
70 & 0.046& 0.169&  20.549& 9.532& 14.712& 11.161\\
 \end{tabular}
 \end{center}
\end{table}

\begin{table}
 \caption{As in Table I, the total suppression factors 
($\cal S$), average energy transfers ($<\!\omega\!>$)
and $rms$ energy transfers ($\omega_{rms}$), but for various temperatures
at a given incident neutrino energy ($E_1$ = 20 MeV).  
The $0$ subscript or superscript corresponds 
to the case without many--body effects, but with final--state nucleon blocking.
This table corresponds to Figure 4, but also includes numbers for
$T$ = 3 MeV.
}
\begin{center}
 \begin{tabular}{ccccccc}
$T$ (MeV)  & $\cal S$& $\cal S_{\it 0}$ & $<\!\omega\!>$ (MeV) & $<\!\omega\!>_0$ (MeV)&
$\omega_{rms}$ (MeV) & $\omega_{rms}^0$ (MeV) \\
\hline
3 & 0.021& 0.093&  2.693&  0.598& 6.623& 4.292\\
5 & 0.038& 0.164& -2.158& -0.970& 9.497& 4.977\\
7 & 0.065& 0.240& -6.153& -1.856& 11.221& 5.358\\
10& 0.118& 0.350& -10.193& -2.699& 12.482& 5.810\\
15& 0.227& 0.516& -14.355& -3.734& 13.928& 6.614\\
20& 0.351& 0.659& -17.572& -4.726& 15.695& 7.618\\
30& 0.639& 0.900& -24.603& -7.118& 21.660& 10.575\\
 \end{tabular}
 \end{center}
\end{table}

\begin{table}
 \caption{As in Tables I \& II, the total suppression factors 
($\cal S$), average energy transfers ($<\!\omega\!>$)
and $rms$ energy transfers ($\omega_{rms}$), but for various mass densities ($\rho$)
at a given incident neutrino energy ($E_1$ = 20 MeV) and for a temperature of 5 MeV.
The $0$ subscript or superscript corresponds 
to the case without many--body effects, but with final--state nucleon blocking.
This table is related to Figure 5.   
}
\begin{center}
 \begin{tabular}{ccccccc}
$\rho$ (gm cm$^{-3}$)  & $\cal S$& $\cal S_{\it 0}$ & $<\!\omega\!>$ (MeV) & $<\!\omega\!>_0$ (MeV)&
$\omega_{rms}$ (MeV) & $\omega_{rms}^0$ (MeV) \\
\hline
$10^{12}$& 0.897& 0.952& -0.295& 0.216& 2.556& 2.157\\
$10^{13}$& 0.560& 0.794& -0.455& 0.256& 3.211& 2.359\\
$3\times10^{13}$& 0.307& 0.591& -0.766& -0.337& 4.269& 2.725\\
$10^{14}$& 0.117& 0.162& -1.525& -0.562& 6.532& 3.598\\
$3\times10^{14}$& 0.038& 0.164& -2.156& -0.970& 9.492& 4.977\\
 \end{tabular}
 \end{center}
\end{table}

\begin{table}
 \caption{As in the other tables, the total suppression factors 
($\cal S$), average energy transfers ($<\!\omega\!>$)
and $rms$ energy transfers ($\omega_{rms}$), but for a post--bounce 
profile from Burrows, Hayes, \& Fryxell (1995),
at a given incident neutrino energy ($E_1$ = 20 MeV).
The $0$ subscript or superscript corresponds 
to the case without many--body effects, but with final--state nucleon blocking.
This table is related to Figure 6.  The density and 
temperature at the edited points in the model
are indicated.                                                                                         
}
\begin{center}
 \begin{tabular}{cccccccc}
$\rho$ (gm cm$^{-3}$)  & $T$ (MeV)& $\cal S$& $\cal S_{\it 0}$ & $<\!\omega\!>$ (MeV) & $<\!\omega\!>_0$ (MeV)&
$\omega_{rms}$ (MeV) & $\omega_{rms}^0$ (MeV) \\
\hline
$3.945\times10^{14}$& 5.053& 0.028& 0.136& -2.142& -1.136& 10.082& 5.426 \\
$3.084\times10^{14}$& 5.518& 0.043& 0.180& -3.305& -1.261&10.135& 5.137 \\
$2.049\times10^{14}$& 9.457& 0.147& 0.402& -7.322& -2.108&10.290& 5.130 \\
$1.281\times10^{14}$& 10.561& 0.243& 0.547& -5.797& -1.882&8.656& 4.706 \\
$6.463\times10^{13}$& 13.531& 0.477& 0.792& -4.629& -1.946&7.460& 4.659 \\
$4.322\times10^{13}$& 14.554& 0.608& 0.880& -4.065& -1.959&6.994& 4.666 \\
$2.669\times10^{13}$& 15.611& 0.750& 0.953& -3.660& -2.013&6.676& 4.735 \\
$5.972\times10^{12}$& 11.287& 0.876& 0.968& -1.740& -1.159&4.554& 3.609 \\
$1.082\times10^{12}$& 6.137& 0.907& 0.957& -0.509& -0.369&2.874& 2.416 \\
 \end{tabular}
 \end{center}
\end{table}

{}

\newpage

\newpage

\vspace*{-1.0in}

\newpage 

\vspace*{-1.0in}

\newpage 

\vspace*{-1.0in}

\newpage

\vspace*{-1.0in}

\newpage

\vspace*{-1.0in}

\newpage

\vspace*{-1.0in}

\newpage

\vspace*{-1.0in}

\newpage

\vspace*{-1.0in}

\newpage

\vspace*{-1.0in}

\newpage

\end{document}